\documentclass[12pt,a4paper, twoside, openright]{book}
\usepackage{amsmath, amssymb, amsthm}
\usepackage{graphicx,color}
\usepackage[left=1.5in, right=1in, top=1in, bottom=1in, includefoot, headheight=13.6pt]{geometry}
\usepackage[square, comma, numbers, sort&compress]{natbib}
\usepackage[T1]{fontenc}
\usepackage[titletoc,page]{appendix}

\usepackage[dvipsnames,svgnames,x11names,hyperref]{xcolor}

\usepackage[hang, small, bf, margin=20pt, tableposition=top]{caption} 
\numberwithin{equation}{section}

\usepackage{fancyhdr}
\makeatletter
\renewcommand{\chaptermark}[1]{\markboth{\textsc{\@chapapp}\ \thechapter:\ #1}{}}
\makeatother

\usepackage{sectsty}
\chapterfont{\large\sc\centering}
\chaptertitlefont{\centering}
\subsubsectionfont{\centering}

\usepackage[colorlinks=false]{hyperref}
\usepackage[figure,table]{hypcap} 
\hypersetup{
    bookmarksnumbered,
    pdfstartview={FitH},
    citebordercolor={red},
    linkbordercolor={blue},
    urlbordercolor={ForestGreen},
    pdfpagemode={UseOutlines}
}

\usepackage{epstopdf}

\usepackage{amsfonts}
\usepackage{slashed} 

\newcommand{\ISpin}{\mathrm{ISpin}}
\newcommand{\Pin}{\mathrm{Pin}}
\newcommand{\Spin}{\mathrm{Spin}}

\newcommand{\SU}{\mathrm{SU}}
\newcommand{\U}{\mathrm{U}}
\newcommand{\OO}{\mathrm{O}}    
\newcommand{\SO}{\mathrm{SO}}
\newcommand{\so}{\mathrm{so}}
\newcommand{\su}{\mathrm{su}}

\newcommand{\dd}{\mathrm{d}}      
\newcommand{\ISO}{{\mathrm{ISO}}}
\newcommand{\tr}{{\mathrm{tr}}}

\newcommand{\R}{\mathbb{R}}  
\newcommand{\C}{\mathbb{C}}  
\newcommand{\Z}{\mathbb{Z}}

\newcommand{\D}{\mathrm{D}} 
\newcommand{\vol}{\mathrm{vol}} 
\newcommand{\diag}{\mathrm{diag}}

\makeatletter

\newcommand{\Rmnum}[1]{\expandafter\@slowromancap\romannumeral #1@}
\makeatother

\numberwithin{equation}{section}

\DeclareMathOperator{\sgn}{sgn}

\DeclareMathOperator{\Realpart}{Re}

\newcommand{\bah}[1]{{\overline{#1}}}

\theoremstyle{definition}
\newtheorem*{definition}{Definition}
\newtheorem*{theorem}{Theorem}

\newtheorem*{proposition}{Proposition}

\begin{document}

\title{Topological quantum field theory and quantum gravity}

\author{STEVEN KERR, MPhys.
\\ \\
Thesis submitted to the University of Nottingham \\
 for the degree of Doctor of Philosophy.\\
\\
}

\maketitle

\newpage \pagestyle{plain}
\vspace*{\fill}

\newpage
\chapter*{Abstract}
This thesis is broadly split into two parts. In the first part, simple state sum models for minimally coupled fermion and scalar fields are constructed on a $1$-manifold. The models are independent of the triangulation and give the same result as the continuum partition functions evaluated using zeta-function regularisation. Some implications for more physical models are discussed.

In the second part, the gauge gravity action is written using a particularly simple matrix technique. The coupling to scalar, fermion and Yang-Mills fields is reviewed, with some small additions. A sum over histories quantisation of the gauge gravity theory in $2+1$ dimensions is then carried out for a particular class of triangulations of the three-sphere. The preliminary stage of the Hamiltonian analysis for the $(3+1)$-dimensional gauge gravity theory is undertaken.

\newpage \pagestyle{plain}
\vspace*{\fill}

\newpage
\chapter*{Acknowledgments}
Primarily I would like to thank my supervisor Professor John Barrett for his excellent tutelage and patience over the course of this PhD.

I would like to thank the department of Mathematical Sciences at the University of Nottingham, all its faculty members, postdocs, graduate students and administrative staff, and in particular, the members of the quantum gravity group. I thank Jorma Louko and Sara Tavares for discussions.

Thank you to my sister, for giving me her physics text books when I was young.

And finally, to my parents; this PhD thesis is as much mine as it is yours.

\newpage \pagestyle{plain}
\vspace*{\fill}

\newpage
\tableofcontents

\listoffigures

\newpage \pagestyle{plain}
\vspace*{\fill}

\newpage
\pagestyle{plain}
\vspace*{\fill}
\begingroup
{\it \noindent `Three passions, simple but overwhelmingly strong, have governed my life: the longing for love, the search for knowledge, and unbearable pity for the suffering of mankind. These passions, like great winds, have blown me hither and thither, in a wayward course, over a great ocean of anguish, reaching to the very verge of despair.
\\
\\
I have sought love, first, because it brings ecstasy - ecstasy so great that I would often have sacrificed all the rest of life for a few hours of this joy. I have sought it next, because it relieves loneliness - that terrible loneliness in which one shivering consciousness looks over the rim of the world into the cold unfathomable lifeless abyss. I have sought it finally, because in the union of love I have seen, in a mystic miniature, the prefiguring vision of the heaven that saints and poets have imagined. This is what I sought, and though it might seem too good for human life, this is what - at last - I have found.
\\
\\
With equal passion I have sought knowledge. I have wished to understand the hearts of men. I have wished to know why the stars shine. And I have tried to apprehend the Pythagorean power by which number holds sway above the flux. A little of this, but not much, I have achieved.
\\
\\
Love and knowledge, so far as they were possible, led upward toward the heavens. But always pity brought me back to earth. Echoes of cries of pain reverberate in my heart. Children in famine, victims tortured by oppressors, helpless old people a burden to their sons, and the whole world of loneliness, poverty, and pain make a mockery of what human life should be. I long to alleviate this evil, but I cannot, and I too suffer.
\\
\\
This has been my life. I have found it worth living, and would gladly live it again if the chance were offered me.'}
 \\
 \\
\endgroup
\vspace*{\fill}
 \quad\quad\quad\quad\quad\quad \quad\quad\quad\quad\quad\quad\quad\quad\quad\quad\quad\quad\quad\quad\quad\quad\quad\quad\quad\quad\quad - Bertrand Russell
 
\newpage \pagestyle{plain}
\vspace*{\fill}

\newpage
\chapter{Introduction}

The problem of quantum gravity has a long history dating back to the 1930's. Since then a number of different approaches have developed, with the two most popular research programmes currently being `string theory' and `loop quantum gravity'. 

There have been many motivations behind this research. One was the discovery in the 1980's that gravity is not perturbatively renormalisable \cite{UV gravity}. Nonetheless quantum gravity does exist as an effective field theory. Indeed the theory may be non-perturbatively renormalisable if there is an asymptotic safety scenario \cite{AS}.

Another motivation is the presence of singularities in general relativity, and the hope that these may be cured by a proper treatment of quantum gravity. However, the singularities of general relativity may well be resolvable at the classical level. For example, the Sciama-Kibble theory of gravity \cite{Kibble, Sciama} may avoid gravitational singularities \cite{Pop}.

Yet another motivation is the unparsimonious, or potentially inconsistent nature of the coupling between a classical gravitational field and quantum matter fields. However, the discussion on this subject has mostly focused on the scenario where the curvature tensor is coupled to the expectation value of the energy-momentum tensor,

\begin{align}
R_{\mu \nu} -\frac{1}{2} g_{\mu \nu} R + g_{\mu \nu} \Lambda = 8\pi \langle \hat{T}_{\mu \nu} \rangle.
\end{align}
Theories of this type face serious, perhaps fatal difficulties. However, there may be other possibilities for a consonant treatment of interacting classical and quantum fields. For example, emergent gravity ala Sakharov \cite{Sakharov} may avoid known difficulties of hybrid quantum/classical systems. 

My conclusion from this brief discussion is that while these are all appealing motivations for research in quantum gravity, there are other approaches that may be viable, and it is important to keep these in mind.

I now very briefly summarise the main modern approaches to quantum gravity.

In the 1980's it was shown that a perturbative quantum field theoretic treatment of the gravitational field about a fixed spacetime background leads to non-renormalisable divergences \cite{UV gravity}. Therefore if there is no asymptotic safety scenario for gravity, this treatment of quantum gravity may at best be considered an effective theory. Since then there have been efforts to find a suitable ultraviolet completion of the theory. This search led to supergravity in the 1980's, and then to superstring theory and `M-theory' in modern times. The driving force behind this line of research has been to find a theory that includes general relativity as a low energy limit and that has a convergent perturbative expansion about a fixed spacetime background.

In 1959, Arnowitt, Deser and Misner developed the ADM formalism \cite{ADM}, which is a Hamiltonian treatment of the Einstein-Hilbert action. However, attempts at canonical quantisation following Dirac's procedure \cite{Lectures on QM} were unsuccessful, mainly due to the complicated form of the Hamiltonian constraint. In 1986, Ashtekar introduced new variables \cite{Ashtekar} that greatly simplified the constraints. This eventually led to the `loop quantum gravity' research programme. The driving force behind this line of research has been a canonical quantisation of gravity in line with Dirac's programme \cite{Lectures on QM}.

There have been numerous other lines of attack on the problem of quantum gravity, notably the sum over histories approach \cite{Baez} which appears to be closely related to loop quantum gravity, and non-commutative geometry \cite{Connes}. 

There have been two main ideas that have guided the research in this thesis. The first is the importance of matter. Much work on quantising gravity in the canonical and sum over histories approaches has focused on quantising the gravitational field in isolation. However, matter is a crucial part of our universe and if there is to be a quantum theory of gravity, it must accommodate matter in a sensible way. Indeed, as Einstein pointed out, the presence of matter is necessary in general relativity for space and time to have any physical meaning at all. Therefore I have attempted to see where the guiding principle of parsimonious matter couplings leads. The result has been the one-dimensional state sum models in part \Rmnum{1} of this thesis.

The other idea has been to take seriously the concept of gravity as a gauge theory. The first order formalism for gravity in $2+1$ dimensions is known to be a gauge theory \cite{Witten} - Chern-Simons theory. In part \Rmnum{2} of this thesis, a gauge theory of gravity in any number of spacetime dimensions is presented. This theory can be coupled to matter in a simple, gauge invariant fashion, and indeed the chronological order of our work was matter coupling to gravity first, gauge gravity second. Thus the two main ideas found an harmonious synthesis.

\part{One-dimensional state sum models}

Traditionally there have broadly been two paths towards quantising a classical theory. The first is the canonical or Hamiltonian formalism, in which one starts out with a phase space that has the structure of a symplectic manifold, and a Hamiltonian function that induces a flow that represents time evolution. Dirac noticed that the mathematical formalism of Hamiltonian dynamics makes manifest certain analogies between classical and quantum systems. Thus there are various procedures for passing from a classical Hamiltonian system to the corresponding quantum system.

The second route is the sum over histories formalism. In this approach, probability amplitudes between an initial and final state are symbolically obtained as a weighted sum over all possible intermediate states. In Lorentzian signature, the weighting factor is $e^{iS}$, where $S$ is the classical action for a particular configuration. In practice, it is most convenient to work with the partition function, 

\begin{align}
{\mathbb Z} = \int \dd \phi \; e^{iS[\phi]}. \label{intro PF}
\end{align}
Here $\phi$ symbolically represents the set of dynamical variables in the theory modulo gauge transformations, with a suitably chosen measure $\dd \phi$. All transition amplitudes may be obtained from \eqref{intro PF} by adding appropriate source terms to the action and then differentiating. 

A mathematically rigorous approach to defining the partition function is via lattice discretisation. This consists of quantising only a finite sample of the original degrees of freedom of the theory. The resulting partition function is often called a `state sum model', because the integrals in \eqref{intro PF} typically reduce to discrete sums.

State sum models are typically constructed from a classical theory on a spacetime manifold $M$ by working with a discrete approximation to $M$, e.g. a triangulation. A special class of state sum models occurs when the partition function is independent of the particular triangulation that one works with. This is an example of a topological quantum field theory (TQFT). In this case, the partition function defines a topological invariant of the manifold $M$. Such theories are of particular interest in the study of quantum gravity because the quantum theory has the same symmetry as the classical theory, namely spacetime diffeomorphism symmetry. 

\begin{figure}[h!]  
\begin{center}
\includegraphics[scale=0.7]{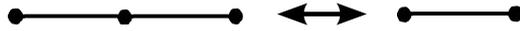}
\end{center} 
\caption{The one-dimensional Pachner move.}\label{1dpachner}
\end{figure}

In this part, simple topological state sum models for fermionic and scalar fields on a one-dimensional manifold will be explored. This provides a simple setting for studying such theories, in no small part because the state sum model must only be invariant under the single one-dimensional Pachner move depicted in figure \ref{1dpachner}. It is hoped that this will help to pave the way for the construction of analogous models in higher dimensions.

\vspace{10mm}

\chapter{A topological state sum model for fermions on the circle} \label{sec: fermions}

In this chapter a simple one-dimensional state sum model is presented in which a fermionic field is coupled to a background gauge field. These results were published in \cite{circle}. A simple formula for the partition function of this model on a triangulated circle (i.e. a polygon) is presented in subsection~\ref{fermion definition}. It is demonstrated that the partition function is independent of the triangulation and depends only on the holonomy of the gauge field. 

In subsection \ref{wkfsection}, it is shown that the state sum model has an action that is a discretisation
of the continuum Dirac action for a massless fermion field coupled to
the gauge connection. Then, in section \ref{fisection} the partition function of the continuum theory is
calculated precisely using zeta function regularisation, to show that it is equal to the result from the state sum model.

The results are extended in subsection \ref{masssection} to models with a mass term. Discretisation of the mass term in a `na\"{i}ve way breaks the triangulation independence of the model. However, it is possible to include a mass term in the model by treating the mass parameter as a gauge field for the appropriate group.

The results presented here complement previous work constructing
quantum gravity state sum models with fermion fields in dimension
three \cite{Fairbairn:2006dn, Dowdall:2010ej} and four
\cite{Bianchi:2010bn, Han:2011as}. These works construct discrete
analogues of the continuum Dirac functional integral according to the
heuristic continuum limit, as considered here in
Subsection~\ref{wkfsection}, but do not have a direct comparison with
the partition function of the continuum functional integral. The
results presented here give the first precise comparison of a discrete
fermionic model with the continuum partition function.

It is an interesting question as to whether these results can be generalised to a higher-dimensional model. Some related properties of the Dirac operator with a gauge field on a graph have been studied previously~\cite{bolte}; this suggests there may also be an extension of the state sum model to graphs.

\section{The state sum model} \label{ssmsection}
In this section, the definition of the fermionic state sum model will be developed on the circle and the interval. First fermionic variables are reviewed. These variables naturally belong to a Grassmann algebra, which is an algebra with an anti-commutative composition law. An integral calculus is developed in the Grassmann setting following \cite{Peskin}, and this is used to define the state sum model.

\subsection{Grassmann algebra}
In order to generate the correct statistics for fermions in quantum field theory, fermionic variables must mutually anti-commute. This notion is captured mathematically by the concept of Grassmann or exterior algebra. The fermions themselves are spinors, which are elements of a vector space with an inner product that is preserved by the spin group $\Spin(p,q)$, the double cover of the special orthogonal group $\SO (p,q)$ whose defining representation is on a vector space with metric signature $(p,q)$. 

\definition Given a vector space $V$, the tensor algebra is defined by
\begin{align}
T(V) = \oplus_{k=0}^{\infty} T^k(V),
\end{align}
where $T^k(V)$ is the $k$-th tensor power of $V$; that is, the vector space obtained by taking the tensor product of $k$ copies of $V$. The product in the algebra is the tensor product.
\\
\\
The Grassmann algebra $\Lambda(V)$ is defined as the algebraic quotient of the tensor algebra $T(V)$ by the two sided ideal $I$ generated by all elements of the form $x \otimes x$, with $x \in V$. Symbolically,

\begin{align}
\Lambda(V) = T(V) / I.
\end{align}
The equivalence class $[x \otimes y]$ is often denoted $x \wedge y$, and indeed $\Lambda(V)$ is an algebra with product given by $\wedge$. This is called the exterior product. In what follows however, we will omit the $\wedge$ and simply write $xy$ for $x \wedge y$.

Alternatively the exterior algebra $\Lambda(V)$ may be constructed as the algebra determined by a number of generators $a_1,a_2,\ldots,a_l$ that form a basis for $V$, subject to the relations
\begin{align}
a_i a_j+a_j a_i=0.
\end{align}
It is clear that $a_i^2=0 \; \forall i$. In quantum field theory, the fermionic variables are Grassmann-valued operators on a suitably defined Hilbert space, and may be expanded in terms of creation and annihilation operators. The relation $a_i^2=0$ captures the idea that it is impossible for two identical fermions to occupy the same quantum state, i.e. the Pauli exclusion principle.

Due to this anti-commutation law, functions on the exterior algebra have a number of curious properties. A function of the generators $f(a_1,\ldots, a_l)$ is a polynomial that terminates at the highest monomial $a_la_{l-1}\ldots a_1$. Because of this, there is no such thing as a transcendental function on the exterior algebra. For example, the exponential function of a single Grassmann variable, which is defined by its power series expansion, terminates after two terms,

\begin{align}
e^x = 1+ x.
\end{align}

In the path integral approach to quantum field theory, one symbolically integrates over the set of all possible field configurations. In the fermionic case this will require us to integrate over a particular Grassmann algebra. For this purpose it is necessary to develop an integral calculus in the Grassmann algebra setting.

The Lebesgue integral over the whole of a given space may be thought of as a linear functional that sends a suitable set of functions to the underlying field. This integral has the property of translation invariance. These two properties, linearity and translation invariance, can be used to axiomatise an integral over the whole of a Grassmann algebra that is called the Berezin integral. That is, we demand that

\begin{align}
\int \dd x (\alpha f(x) + \beta g(x)) &= \alpha \int  \dd x f(x) + \beta \int \dd x g(x), \\
\int \dd x (x+y) &= \int \dd x \; x,
\end{align}
where $\alpha$, $\beta$ are scalars in the underlying field. The latter condition implies that $\int \dd x = 0$. The integral will be normalised so that $\int \dd x \; x = 1$. For iterated integrals,  

\begin{align}
\int\dd a_1\,\dd a_2\ldots\dd a_l\; f(a_1,\ldots, a_l)
\end{align}
is defined to be the coefficient of $a_la_{l-1}\ldots a_1$ in the expansion of $f$. No independent meaning is attached to the differentials in these formulae, and they do not appear outside an integral. However the order of them in the integral is important; transposing two neighbouring differentials in the notation changes the sign of the integral.

A curious property of the Berezin integral is its behaviour under a change of variables. In direct contrast to the usual change of variables formula, the measure of the Berezin integral transforms with an {\it inverse} factor of the Jacobian determinant,

\begin{align}
&\int\dd a_1\,\dd a_2\ldots\dd a_l\; f(a_1,\ldots, a_l) = \nonumber \\ 
\int (\det J)^{-1} &\dd a'_1\,\dd a'_2\ldots\dd a'_l\; f(a_1(a'_1 \ldots a'_l ),\ldots, a_l(a'_1 \ldots a'_l )),
\end{align}
where $J$ is the Jacobian for the change of variables $a_i \rightarrow a_i'$. This property follows as a direct consequence of the definition of the Berezin integral, and is proved in \cite{Peskin}.

It is possible to extend the above definitions to integration over a subset of coordinates, and perform the integral iteratively. So if $f=a_ka_{k-1}\ldots a_1 b$, where $b$ is a polynomial in the remaining variables $a_{k+1},\ldots, a_l$, then the integral is
\begin{align}
\int\dd a_1\,\dd a_2\ldots\dd a_k\; f= b,
\end{align}
with terms multiplying lower degree monomials in $a_1\ldots a_k$ integrating to zero. An example of iteration is the formula
\begin{align}
\int\dd a_1\,\dd a_2\; f=\int\dd a_1\,\left(\int\dd a_2\; f\right).
\end{align}

In the applications of interest here, the generators of the Grassmann algebra occur in pairs $a_i$, $b_i$ that form the components of $n$-dimensional vectors, 
\begin{align}
\psi=(a_1,a_2,\ldots a_n),\quad \overline{\psi}=(b_1,b_2,\ldots,b_n).
\end{align}
In this case the integral is defined with the notation
\begin{align}
\int \dd \psi\, \dd \overline{\psi} = \int \dd a_1\, \dd b_1\, \dd a_2\, \dd b_2 \ldots \dd a_n\, \dd b_n.
\end{align} 

Let $M$ be an $n\times n$ matrix with entries in $\C$. The Grassmann analogue of gaussian integration is
\begin{align}
\int \dd\psi\, \dd\overline{\psi} \, e^{\overline{\psi} M \psi}   = \det M,  
\label{basicgaussian}  
\end{align}
which is proved by expanding the exponential. 

The result \eqref{basicgaussian} can be extended to the case of fermionic
source terms. Take $\overline{c}, d$ to be $n$-component vectors with
Grassmann-valued entries that are polynomial of odd degree in the
remaining generators (i.e. excluding components of $\psi$ and
$\overline{\psi}$ respectively), and $M$ now an invertible matrix. 
We then have
\begin{align}  
\int \dd\psi\, \dd\overline{\psi} \, 
e^{\overline{\psi} M \psi + \overline{c} \psi + \overline{\psi} d} 
= \det M \, e^{-\overline{c} M^{-1} d}, 
\label{gaussian}
\end{align}
which is proved by first completing the square with the translations 

\begin{align}
\overline{\psi}&\mapsto \overline{\psi} - \overline{c} M^{-1}, \\ 
\psi&\mapsto \psi -  M^{-1}d,
\end{align}
and then using~\eqref{basicgaussian}.

\subsection{Definition of the state sum model} \label{fermion definition}

\begin{figure}[h!]  
\begin{center}
\includegraphics[scale=0.7]{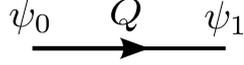}
\end{center} 
\caption{The fermionic state sum model on an edge. The arrow indicates the orientation of the edge.}\label{fig1}
\end{figure}

Start with an oriented interval $[0,l]$ of length $l$, triangulated with $N+1$ vertices. The vertices are labelled by $i=0...N$, and each is decorated with $n$-dimensional vectors $\psi_i$, $\overline{\psi}_i$. The edges are decorated with invertible $n \times n$ matrices $Q_{i,i+1}$ satisfying $Q_{i,i+1} = Q_{i+1,i}^{-1}$. We define $Q= \prod_{i=0}^{N-1} Q_{i,i+1}$. The relevant Grassmann algebra is the one generated by all of the components of all of the vectors. For $N=1$, i.e. a single edge, the state sum model is
 
\begin{align}
{{\mathbb Z}}^Q_{[0,l]}= e^{-\overline{\psi}_0 Q \psi_1}. \label{interval}
\end{align}
This has fermionic variables $\overline{\psi}_0$ and $\psi_1$ associated to vertices $0$ and $1$ respectively, as depicted in figure \ref{fig1}.

Gluing two such partition functions together is carried out using the following proposition, which states that one can multiply matrices by the use of Berezin integration.
\proposition 
\begin{align}
\int \dd\psi_1 \dd\overline{\psi}_1 e^{-\overline{\psi}_0 Q_{0,1} \psi_1} e^{\overline{\psi}_1 \psi_1} e^{-\overline{\psi}_1 Q_{1,2} \psi_2} = e^{-\overline{\psi}_0 Q_{0,1} Q_{1,2} \psi_2}. \label{proposition1}
\end{align}

\begin{proof}
We have 
\begin{align}
\int \dd\psi_1 \dd\overline{\psi}_1\, e^{-\overline{\psi}_0 Q_{0,1} \psi_1} e^{\overline{\psi}_1 \psi_1} e^{-\overline{\psi}_1 Q_{1,2} \psi_2} &= \int \dd\psi_1 \dd\overline{\psi}_1 e^{\overline{\psi}_1 I \psi_1 - (\overline{\psi}_0 Q_{0,1})\psi_1 - \overline{\psi}_1 (Q_{1,2} \psi_2)},
\end{align}
where $I$ is the $n\times n$ identity matrix. 
The result \eqref{proposition1} follows using 
\eqref{gaussian} with
\begin{align}
\overline{c}=-\overline{\psi}_0 Q_{0,1}
, \ \ 
d=-Q_{1,2} \psi_2
, \ \ 
M=I.
\end{align}
\end{proof}
The formula \eqref{proposition1} is interpreted as a bilinear form on the fermionic states,
\begin{align}
(f,g)=\int \dd\psi_1 \dd\overline{\psi}_1\, f(\psi_1)\, e^{\overline{\psi}_1 \psi_1} g(\overline{\psi}_1),  
\end{align}
and using this bilinear form to glue the partition functions results in
\begin{align}
\label{glueing}
( {{\mathbb Z}}^{Q_{0,1}}_{[0,l]}, {{\mathbb Z}}^{Q_{1,2}}_{[0,l]} )={{\mathbb Z}}^Q_{[0,l]},
\end{align}
with $Q=Q_{0,1}Q_{1,2}$.

This procedure can be iterated for the multiplication of 
any finite number of matrices, yielding
the definition of the state sum model on the interval. Explicitly, 
\begin{align}
\int \dd\psi_1\, \dd\overline{\psi}_1 
\ldots \dd\psi_{N-1}\, &\dd\overline{\psi}_{N-1} \; e^{-\overline{\psi}_0 Q_{0,1} \psi_1} e^{\overline{\psi}_1 \psi_1} 
e^{-\overline{\psi}_1 Q_{1,2} \psi_2} 
\ldots 
e^{-\overline{\psi}_{N-1} Q_{N-1,N} \psi_N} \notag \\
&= e^{-\overline{\psi}_0Q\psi_N}={{\mathbb Z}}^Q_{[0,l]},
\label{repeated Berezin integral} 
\end{align}
with $Q=Q_{0,1}Q_{1,2}\ldots Q_{N-1,N}$.

The leftmost expression in \eqref{repeated Berezin integral} 
is to be interpreted as the definition of the fermionic state sum model on an interval that is triangulated using $(N+1)$ vertices. The state sum model is triangulation independent; indeed, \eqref{glueing} is a statement of the independence of the state sum model under the one-dimensional Pachner move \ref{1dpachner}. It can be modified to include observables, that is, non-trivial functions of the intermediate variables $\psi_i$, $\overline{\psi}_i$. In this sense the model is richer than the evaluation of the partition function on the right-hand side. 

\begin{figure}[h!]  
\begin{center}
\includegraphics[scale=0.7]{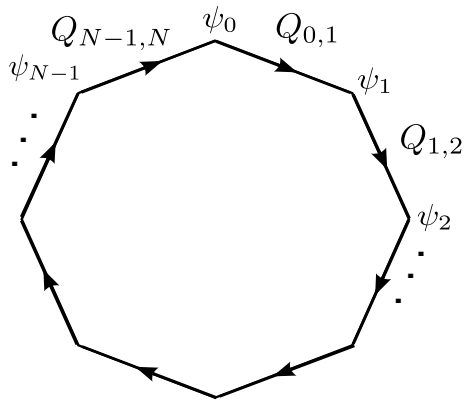}
\end{center} 
\caption{The fermionic state sum model on a triangulated circle.}\label{fig2}
\end{figure}

The partition function for the circle can be computed by gluing together the endpoints of the interval, as depicted in figure \ref{fig2}.

Mathematically, this is done by including an extra factor
$e^{\overline{\psi}_N \psi_N}$ in the integrand, identifying
$\psi_0=\psi_N$, $\overline{\psi}_0=\overline{\psi}_N$, and
integrating over the newly introduced variables,
\begin{align}
{{\mathbb Z}}_{\mathrm{S}^1}^Q
&=  \int \dd\psi_N \dd\overline{\psi}_N\,  
e^{\overline{\psi}_N \psi_N} e^{-\overline{\psi}_N  Q\psi_N} 
= \det \left(  I - Q\right), 
\label{circle} 
\end{align}
where the last equality follows from 
\eqref{basicgaussian} and the observation that $\overline{\psi}_N \psi_N$ and 
$\overline{\psi}_N  Q\psi_N$ commute. 

An immediate consequence of \eqref{circle} 
is that ${{\mathbb Z}}_{\mathrm{S}^1}^Q$ 
vanishes if $Q$ has an eigenvalue equal to~$1$, 
as is for example the case if $Q \in \SO(2n+1)$. 
For $Q \in \SO(2n)$, ${{\mathbb Z}}_{\mathrm{S}^1}^Q$ is real-valued,
as the eigenvalues occur in complex conjugate pairs. 
There are however groups for which 
${{\mathbb Z}}_{\mathrm{S}^1}^Q$ is complex-valued. 
An example of particular interest is~$\U(1)$, for which we have 
\begin{align}
\U(1) : \quad 
{{\mathbb Z}}_{\mathrm{S}^1}^Q= 1 - Q, 
\quad Q = e^{-i\theta}, 
\quad 
\theta\in[0,2\pi). 
\label{UoneZ}
\end{align}

We note that the partition function may be independent of the orientation for some matrices $Q$. An orientation reversal acts by sending $Q$ to $Q^{-1}$. Hence the partition function is independent of the orientation iff $\det \left( I - Q\right) = \det \left( I - Q^{-1}\right)$. For example, in the case where $Q \in \U(1)$, we see from \eqref{UoneZ} that orientation reversal acts on the partition function non-trivially, by complex conjugation.

\subsection{Interpretation of the state sum model}
\label{wkfsection}

The state sum model \eqref{circle} may be interpreted as a discrete path integral quantisation of a minimally coupled fermionic field on the circle.

The $\psi_i$ and $\overline{\psi}_i$ variables are interpreted as a discrete sampling of continuous fermionic fields $\psi(t)$, $\overline{\psi}(t)$ on the circle, with $t\in [0,l)$, so that

\begin{align}
\psi_j = \psi\left(j\Delta t \right), \quad \quad \overline{\psi}_j = \overline{\psi}\left(j\Delta t \right),
\end{align}
where $\Delta t = \frac{l}{N}$ is the length of one edge. Thus, as the notation suggests, $\psi_i$ and $\overline{\psi}_i$ are multiplets of one-dimensional spinors that are complex conjugates of one another. The edge connecting vertices $i$ and $i+1$ is assigned the matrix $Q_{i,i+1}$, which is interpreted as the parallel transporter for the $\psi$ field along that edge. Then $Q_{j,j+1} = e^{-i\int_{j\Delta t}^{(j+1)\Delta t} \dd t  A(t)} $, with $A(t)$ the gauge field on the circle.

The state sum model on a circle triangulated with $N$ edges is
\begin{align}
 {{\mathbb Z}}_{\mathrm{S}^1}^Q &= \int \left( \prod_{j=1}^{N} \dd\psi_j\, \dd\overline{\psi}_j \right) e^{-\overline{\psi}_0 Q_{0,1} \psi_1} e^{\overline{\psi}_1 \psi_1} ...  e^{\overline{\psi}_{N-1} \psi_{N-1}} e^{-\overline{\psi}_{N-1} Q_{N-1,N} \psi_{N}}e^{\overline{\psi}_N \psi_N} \nonumber \\
 &= \int \left( \prod_{j=1}^{N} \dd\psi_j\, \dd\overline{\psi}_j \right) \; e^{i\hat{S}},
\label{circlessm} 
\end{align}
where
\begin{align}
\hat{S} &=  -i\left( -\overline{\psi}_0 Q_{0,1} \psi_1 + \overline{\psi}_1 \psi_1 -\overline{\psi}_1 Q_{1,2} \psi_2 + \ldots \overline{\psi}_{N-1} \psi_{N-1} -\overline{\psi}_{N-1} Q_{N-1,N} \psi_N + \overline{\psi}_N \psi_N\right)
\notag 
\\
&=  i \Delta t \sum_{j=0}^{N-1} \overline{\psi}_j   \left(  \frac{Q_{j,j+1} \psi_{j+1}-  \psi_j}{\Delta t}\right)  ,
\end{align}
and $\psi_{N} =\psi_0$, $\overline{\psi}_{N} =\overline{\psi}_0$. 

Now consider the limit $\Delta t \rightarrow 0$ while keeping $l$ constant. Assuming the field values are differentiable,
\begin{align}
\label{derivative}
\lim_{\Delta t \rightarrow 0} i \left(  \frac{Q_{i,i+1} \psi_{i+1}-  \psi_i}{\Delta t}   \right) 
&=  \slashed{\D} \psi(t), 
\end{align}
where the gauge covariant Dirac operator $\slashed{\D}$ is given by 
\begin{align}
\slashed{\D}=i \frac{\dd}{\dd t} + A(t) . 
\label{diracoper-gen}
\end{align}
The single gamma matrix is equal to the complex number~$i$. 
There is no spin connection contribution to \eqref{diracoper-gen}
because the Lie algebra $\mathfrak{so}(1)$ is trivial.  

In the limit $\Delta t \rightarrow 0$ the sum converges to an integral, 
\begin{align} 
\label{integral}
\lim_{\Delta t \rightarrow 0} \Delta t \sum_{j=0}^{N-1}f(j\Delta t) 
= \int_0^l f(t)\;\dd t ,
\end{align}
and therefore
\begin{align}
\lim_{\Delta t \rightarrow 0} \hat{S}
\, = \, 
 \int_0^l \dd t\; \overline{\psi}(t)  \slashed{\D} \psi(t).
\end{align}
This is precisely the Dirac action for a minimally coupled scalar field on the circle. Thus the state sum model of the previous subsection may be considered a definition of the partition function of the Dirac theory via a lattice discretisation procedure.

Gauge transformations act on the partition function of $[0,1]$ by a linear transformation acting on each set of fermionic variables. Thus if $U_i$, $U_j$ are invertible matrices then the transformation is $\psi_j\to U_j\psi_j$ and $\overline{\psi}_i\to \overline\psi_i U_i^{-1}$. The matrices $Q_{i,i+1}$ transform under $Q'_{i,i+1} =  U_i Q_{i,i+1} U_{i+1}^{-1}$. It is clear then that the partition function for the interval \eqref{interval} and the circle \eqref{circle} are invariant under gauge transformations.

\section{Comparison with the functional integral}
\label{fisection}

In this section, the partition function for minimally coupled fermions fields on the circle is compared to the corresponding functional integral. First zeta function regularisation is briefly reviewed in the context of quantum field theory following \cite{baer-schopka}. This allows an exact evaluation of the Dirac functional integral, which gives a result that is identical to the state sum model.

\subsection{Zeta-function regularisation}
\label{zetasection}
Zeta function regularisation is a method of regularising divergent products and sums. It can be used to regularise certain quantum field theories by giving rigorous meaning to the determinant of operators on infinite-dimensional Hilbert spaces. Here we develop the zeta-function regularised definition of $\det
\D$ for a self-adjoint operator $\D$ that is positive definite following \cite{baer-schopka}, and then generalise this definition to~$\det (i\D)$.

\subsubsection{Positive definite $D$}

To begin, suppose $D$ is a Hermitian, strictly positive operator in a finite-dimensional Hilbert space.  The zeta-function $\zeta_{\D} (s)$ of $\D$ is defined for $s\in\C$ by
\begin{align}
\zeta_{\D} (s) = \sum_k \frac{1}{\lambda_k^s}, \label{eq:zetaLdef}
\end{align}
where $\lambda_k$ are the eigenvalues of~$\D$. As $\D$ has a finite number of positive eigenvalues,  $\zeta_{\D} (s)$ is well defined and holomorphic in~$s$. We have

\begin{align}
\det \D = \prod_k \lambda_k = \prod_k e^{\ln \lambda_k} = e^{\sum_k \ln \lambda_k }.
\end{align}
Now consider

\begin{align}
&-\frac{\dd}{\dd s} \lambda_k^{-s} = -\frac{\dd}{\dd s} e^{-s \ln \lambda_k} = \ln \lambda_k e^{-s \ln \lambda_k} \notag \\
&\quad \quad \quad \Rightarrow \ln \lambda_k = \left(-\frac{\dd}{\dd s} \frac{1}{\lambda_k^s} \right) \bigg|_{s=0} \label{eq2}.
\end{align}
Therefore we have

\begin{align}
\det \D = e^{\sum_k \left( -\frac{\dd}{\dd s} \frac{1}{\lambda_k^s} \right) \Big|_{s=0} } =  e^{\left(-\frac{\dd}{\dd s} \sum_k   \frac{1}{\lambda_k^s} \right) \Big|_{s=0} } = e^{- \zeta_{\D'(0)} }. \label{eq:detLdef}
\end{align}
The point of \eqref{eq:detLdef} is that the expression on the right hand side may, under certain circumstances, be taken as a definition of $\det \D$ even when the Hilbert space is infinite-dimensional. We require the spectrum of $\D$ to be discrete, and the sum in \eqref{eq:zetaLdef} must converge for sufficiently large $\Realpart s$ to define $\zeta_{\D} (s)$ as a function that can be analytically continued to $s=0$.  The analytic continuation in $s$ provides a prescription for regularising the divergent product~$\prod_k \lambda_k$.
\subsubsection{Indefinite $\D$}

Now suppose that $\D$ is an indefinite Hermitian operator in a
finite-dimensional Hilbert space, such that the spectrum of $\D$ does
not contain zero.  We wish to express $\det(i\D)$ in a
form similar to\eqref{eq:detLdef}.  The new issue is to accommodate the negative and imaginary eigenvalues. 

Let $\lambda_k$ denote the eigenvalues of~$\D$, enumerated so that
$\lambda_k>0$ for $k>0$ and $\lambda_k<0$ for $k\le0$.  
We define two zeta-functions by

\begin{align}
\zeta_{\D,\epsilon} (s) 
&= \sum_{k>0} \frac{1}{\lambda_k^s}
+ e^{i\epsilon\pi s}\sum_{k\le0} \frac{1}{{(-\lambda_k)}^s} , 
\label{eq:zetasDeps-def}
\\
\zeta_{i\D} (s) 
&= e^{-i\pi s/2}\sum_{k>0} \frac{1}{\lambda_k^s}
+ e^{i\pi s/2}\sum_{k\le0} \frac{1}{{(-\lambda_k)}^s}, 
\label{eq:zetasiDeps-def}
\end{align}
where $\epsilon \in \{1,-1\}$. 
For integer argument these functions agree with na\"\i{}vely 
allowing negative or imaginary eigenvalues in~\eqref{eq:zetaLdef}. 

The eta-function of $\D$ is defined by 
\begin{align}
\eta_{\D} (s) 
& = \sum_k \frac{ \sgn \lambda_k}{ {| \lambda_k |}^s}
= \sum_{k>0} \frac{1}{\lambda_k^s} - \sum_{k\le0}
\frac{1}{{(-\lambda_k)}^s} . 
\label{eq:etainv-def}
\end{align}
Finally, since 
$\D^2$ is positive definite, its zeta-function is
defined by the replacements 
$\D\to \D^2$ and $\lambda_k \to \lambda_k^2$
in~\eqref{eq:zetaLdef}. It follows that 
\begin{align}
\zeta_{\D^2} (s/2) 
&= \sum_k \frac{1}{{|\lambda_k|}^s}
= \sum_{k>0} \frac{1}{\lambda_k^s} + \sum_{k\le0} \frac{1}{{(-\lambda_k)}^s} . 
\label{eq:zetasD2}
\end{align}

The functions \eqref{eq:zetasDeps-def}, \eqref{eq:zetasiDeps-def}, \eqref{eq:etainv-def}
and \eqref{eq:zetasD2} are well defined and holomorphic in $s$. They satisfy

\label{eq:threezetas-2nd}
\begin{align}
\zeta_{\D,\epsilon}(s) 
& = \frac12 \! \left(1 + e^{i\epsilon\pi s} \right) \zeta_{\D^2} (s/2) 
+ \frac12 \! \left(1 - e^{i\epsilon\pi s} \right) \eta_{\D} (s) , 
\label{eq:zetasDeps-2nd}
\\[1ex]
\zeta_{i\D} (s) 
& = \cos(\pi s/2) \zeta_{\D^2} (s/2) 
- i \sin(\pi s/2) \eta_{\D} (s) , 
\label{eq:zetasD-2nd}
\end{align}
and differentiation at $s=0$ yields 

\label{eq:threezetas-diff}
\begin{align}
\zeta'_{\D,\epsilon} (0) 
& = \frac12 \zeta'_{\D^2} (0) 
+ \frac{i\epsilon\pi}{2} 
\bigl( 
\zeta_{\D^2} (0) -  \eta_{\D} (0) \bigr) , 
\\[1ex]
\zeta'_{i\D} (0) 
& = \frac12 \zeta'_{\D^2} (0) 
- \frac{i\pi}{2} \eta_{\D} (0) . 
\end{align}

We are now ready to turn to the determinants. 
They are

\begin{align}
\det \D &= \prod_k \lambda_k = e^{- \zeta_{\D,\epsilon}'(0) } = e^{ i\epsilon \frac{\pi}{2} \left(\eta_{\D} (0) - \zeta_{\D^2} (0) \right)} 
\, e^{ - \frac{1}{2} \zeta_{\D^2}' (0)} , 
\label{eq:detsDepsfinal}  
\\
\det(i\D) &= \prod_k (i\lambda_k) = e^{- \zeta_{i\D}'(0) } = e^{ i \frac{\pi}{2} \eta_{\D} (0) } \, e^{ - \frac{1}{2} \zeta_{\D^2}' (0)} . 
\label{eq:detisDfinal} 
\end{align}

Formulas \eqref{eq:detsDepsfinal} and \eqref{eq:detisDfinal} provide definitions for
$\det \D$ and $\det(i\D)$ that extend to the case when the Hilbert space is
infinite-dimensional and separable, and the spectrum of $\D$ is discrete
with suitable asymptotic properties so that the zeta functions \eqref{eq:zetasDeps-def} and \eqref{eq:zetasiDeps-def} are well defined for sufficiently large $s$ and then analytically continued to $s=0$ . The functions $\eta_{\D} (s)$ and
$\zeta_{\D^2} (s/2)$ are 
defined by \eqref{eq:etainv-def}
and \eqref{eq:zetasD2} for sufficiently large $\Realpart s$ and
analytically continued to $s=0$.

An important difference between $\det \D$ and $\det(i\D)$ arises from
the phases in \eqref{eq:zetasDeps-def} and \eqref{eq:zetasiDeps-def}.  In the definition
of $\zeta_{i\D}$~\eqref{eq:zetasiDeps-def}, the phases of the positive and
negative eigenvalue terms were chosen to be opposite for real argument, with
the consequence that in the finite-dimensional case $\zeta_{i\D}$ is
real-valued for real argument whenever the spectrum of $\D$ is
invariant under $\D \to -\D$. In the definition of
$\zeta_{\D,\epsilon}$~\eqref{eq:zetasDeps-def}, by contrast, 
the branch of ${(-1)}^{-s}$ in the negative eigenvalue terms
cannot be fixed by a similar symmetry argument, and the ambiguity was parameterised by $\epsilon \in \{1,-1\}$, which still survives in the final formula 
\eqref{eq:detsDepsfinal} for~$\det \D$. 
In the finite-dimensional case, $\eta_{\D} (0) - \zeta_{\D^2} (0)$ is
an even integer and the right-hand side of \eqref{eq:detsDepsfinal}
is thus independent of~$\epsilon$. In the infinite-dimensional case,
however, the two values of $\epsilon$ can yield different regularised
values for~$\det \D$. The choice $\epsilon=1$ is related
to our regularisation of $\det(i\D)$ since $\zeta_{\D,1}(s) = e^{i\pi
  s/2} \zeta_{i\D}(s)$, whereas the formulas in \cite{baer-schopka}
make the choice $\epsilon=-1$.


\subsection{Dirac determinant on the circle for~$\U(1)$}
\label{sec:evaluation}

In this subsection $\det(\slashed{\D})$ and
$\det(i\slashed{\D})$ are evaluated for the Dirac operator $\slashed{\D}$
\eqref{diracoper-gen} with the gauge group~$\U(1)$, using the
regularisations \eqref{eq:detsDepsfinal}, \eqref{eq:detisDfinal}.

For $\U(1)$, the Dirac operator \eqref{diracoper-gen} reduces to
$\slashed{\D} = i\frac{\dd}{\dd t} + A(t)$, where $A$ is a real-valued
function of the coordinate $t\in[0,l]$, and both $A$ and the domain of $\slashed{\D}$
have periodic boundary conditions. By a gauge transformation, $A$ may be taken to a constant value that will be denoted by $2\pi a/l$,
with $a\in[0,1)$.  The
holonomy of $A$ is $Q = e^{-2\pi i a}$. Note that $a$ is uniquely
determined by the holonomy.

The eigenvectors of $\slashed{\D}$ are the solutions of

\begin{align}
\slashed{\D} \psi (t) = i\frac{\dd \psi(t)}{\dd t} + A\psi(t) = \lambda \psi(t),
\end{align}
subject to the boundary condition $\psi(l) = \psi(0)$. The opposite spin structure may be taken both here and in the state sum model \eqref{circle}, but for definiteness we have made a choice. The conclusions we reach also hold for anti-periodic boundary conditions on the fermions. This has a linearly independent set of solutions given by $\psi=e^{-i \frac{2\pi kt}{l}}$, $k \in \Z$. Substituting these solutions, the eigenvalues are

\begin{align} 
\lambda_k = 2\pi(k+a)/l.
\end{align}
We exclude the special case $a=0$, in which one eigenvalue vanishes.  We then have
$a\in(0,1)$, all the eigenvalues are non-vanishing, and we are in the
situation covered by the previous subsection.

Now we wish to calculate $\zeta_{\slashed{\D}^2}' (0)$. Firstly, $\zeta_{\slashed{\D}^2} (0)$ may be expressed in terms of the Hurwitz zeta function,

\begin{align}
\zeta_{\slashed{\D}^2} (s/2) &= 
{(2\pi/l)}^{-s}
\sum_{k \in \Z} \frac{1}{{|k+a|}^{s}} 
\notag 
\\
&= 
{(2\pi/l)}^{-s}
\left(\sum_{j=0}^{\infty} \frac{1}{{(j+a)}^{s}} 
+ \sum_{j=0}^{\infty} \frac{1}{{(j+1-a)}^{s}} \right)
\notag 
\\
&= 
{(2\pi/l)}^{-s}
\bigl( \zeta_H (s, a) + \zeta_H (s, 1-a) \bigr), \label{more zeta crap}
\end{align}
where the sums are absolutely convergent and the 
Hurwitz zeta function $\zeta_H$ is defined by \cite{dlmf}
\begin{align}
\zeta_H (s,q) = \sum_{j=0}^{\infty} \frac{1}{{(j+q)}^s}.
\end{align}
Analytically continuing to $s=0$ and using 25.11.18 
in~\cite{dlmf},
\begin{align}
&\frac{\dd}{\dd s} \zeta_H (s,q) |_{s=0} =  \ln \Gamma (q) - \frac{1}{2} \ln 2\pi, \label{zeta stuff}
\end{align}
we obtain

\begin{align}
\zeta_{\slashed{\D}^2}' (0) &= 2 \ln \Gamma (a) + 2 \ln \Gamma (1-a) - 2 \ln 2\pi.
\end{align}
Therefore,

\begin{align}
e^{ - \frac{1}{2} \zeta_{\slashed{\D}^2}' (0)} &= e^{-\ln (\Gamma (a) \Gamma(1-a)) + \ln 2\pi} \nonumber \\
&= \frac{2\pi}{\Gamma (a) \Gamma(1-a)} \nonumber \\
&= 2 \sin \pi a. \label{stuff}
\end{align}

Now consider $\eta_{\slashed{\D}}$. Assuming again $\Realpart s > 1$,
\begin{align} 
\eta_{\slashed{\D}} (s) &= 
{(2\pi/l)}^{-s}
\sum_k \frac{ \sgn (k+a)}{ {| k+a |}^s} 
\notag 
\\
&=
{(2\pi/l)}^{-s}
\left(
\sum_{j=0}^{\infty} \frac{1}{{(j+a)}^s} - \sum_{j=0}^{\infty} \frac{1}{{(j+1-a)}^s} 
\right)
\notag 
\\
&= 
{(2\pi/l)}^{-s}
\bigl(
\zeta_H (s,a) - \zeta_H(s,1-a) \bigr).
\end{align}
Analytically continuing to $s=0$ and using 25.11.13 in~\cite{dlmf}, 

\begin{align}
\zeta_H (0,q) = \frac{1}{2} - q \label{Hurwitz limit},
\end{align}
we find 
\begin{align}
\eta_{\slashed{\D}} (0) 
= 1-2a , 
\label{eq:appetainvfinal}
\end{align}
in agreement with \cite[\S 1.13]{gilkey}. Combining \eqref{more zeta crap} and \eqref{Hurwitz limit}, we also have 

\begin{align}
\zeta_{\slashed{\D}^2}(0)=0. \label{zeta shit}
\end{align}
Finally, using \eqref{eq:detsDepsfinal} and \eqref{eq:detisDfinal} with \eqref{stuff}
,~\eqref{eq:appetainvfinal} and \eqref{zeta shit}, and recalling
$Q=e^{-2\pi i a}$,

\begin{align}
\det \slashed{\D}
&= 
\begin{cases}
1 - Q& \text{for $\epsilon=1$},\\
1 - Q^{-1}& \text{for $\epsilon=-1$}, 
\end{cases}
\label{eq:app:final-det-eps}
\\[1ex]
\det(i\slashed{\D})
& = 
1 - Q . 
\label{eq:app:final-det}
\end{align}
Note that $\det \slashed{\D}$ and $\det(i\slashed{\D})$ depend only on
the holonomy and not on~$l$.

The modulus of the final result \eqref{eq:app:final-det} for
$\det(i\slashed{\D})$ agrees with the calculation in the physics
literature of the ratio of two such determinants with different values
of $a$~\cite{paper with mistake}.  However the phase does not agree,
presumably due to the fact that the definition of this ratio in
\cite{paper with mistake} is given as an infinite product that is not
absolutely convergent.

\subsection{The functional integral}
\label{sectionfuncint}

The continuum partition function is given by
\begin{align} 
{\mathbb F}^A_{S^1} = \int \mathcal{D} \psi \mathcal{D} \overline{\psi}  \, 
e^{i\int_0^{l} \dd t \, \overline{\psi}(t) \slashed{\D} \psi(t)} , 
\label{Dirac generating functional} 
\end{align}
where $\slashed{\D}$ is given by \eqref{diracoper-gen} 
and the fermions obey periodic boundary conditions. 
The functional integral in \eqref{Dirac generating functional} is defined 
by the zeta-function
regularisation~\cite{baer-schopka,vassilevich-manual} given in the previous subsection,
\begin{align}
{\mathbb F}^A_{S^1} = \det (i\slashed{\D}) =1-Q.
\label{determinant definition} 
\end{align}
This is identical to the result \eqref{circle} from the state sum for the group $\U(1)$, 
\begin{align}
{\mathbb F}^A_{S^1}={{\mathbb Z}}_{\mathrm{S}^1}^Q,
\label{surprising}
\end{align} 
where $Q$ is the holonomy of the connection~$A$. The result
\eqref{surprising} generalises immediately to $\U(n)$ by diagonalising
the connection with a gauge transformation, whereupon the functional
integral is the product of a number of $\U(1)$ functional
integrals. The Dirac functional integral is invariant under these
gauge transformations because it depends only on the eigenvalues of
the Dirac operator, which are gauge invariant.

The result \eqref{surprising} is surprising because the eigenvalues of
the Dirac operator are unbounded and so the na\"\i{}ve determinant of the
Dirac operator, the product of its eigenvalues, diverges. Somehow the
discrete model both approximates the eigenvalues of the continuum
operator yet also avoids the divergence, and miraculously imitates the
zeta-function regularisation.

Some insight into the result \eqref{surprising} can be gained by
comparing the eigenvalues for the continuum Dirac operator with the
eigenvalues for its discrete version.

The discrete version of the Dirac operator is a matrix $M$ acting on
the vectors $\psi = \bigoplus_{j=1}^{N} \psi_j$. It is determined by
\begin{align}
\overline{\psi} iM \psi= \sum_{j=0}^{N-1} \overline{\psi}_j  
\left(  \psi_j -  Q_{j,j+1} \psi_{j+1} \right) ,
\end{align}
and can be written in block form as 
\begin{align} 
iM =
\left( \begin{array}{ccccc}
 1 & -Q_{1,2} & & & \\
 & 1 & -Q_{2,3} & &  \\
 & & \ddots & \ddots & \\
 & & & 1 & -Q_{N-1,N} \\
 -Q_{0,1} & & & & 1 \end{array} \right),
\end{align}
where each $Q_{i,i+1}$ is an element of $\U(n)$. Since we have seen that the partition
function associated with $M$ is exactly the same as the continuum partition function,
it must be that the matrix $M$ is in some sense approximating the
differential operator~$\slashed{D}$. We now make this more explicit.

For concreteness, we specialise to~$\U(1)$. The 
eigenvalues of $iM$ are $\mu_k =
1-\alpha_k$, where $\alpha_k$ are the $N$ roots of  
\begin{align}
Q=\alpha^N  , 
\end{align}
and the corresponding eigenvectors are 
\begin{align} 
\left( \begin{array}{ccccc}
Q^{-1}_{0,1} \\
\alpha_k Q^{-1}_{1,2} Q^{-1}_{0,1}  \\
\alpha_k^2 Q^{-1}_{2,3} Q^{-1}_{1,2} Q^{-1}_{0,1}\\
\vdots \\
\alpha_k^{N-1} Q^{-1}_{N-1,N}\ldots Q^{-1}_{0,1} 
\end{array} \right).
\end{align}
Taking $Q=e^{-i\theta}$ with $\theta \in [0,2\pi)$ as in~\eqref{UoneZ}, 
\begin{align}
\mu_k = 1-e^{-i\left( \frac{\theta +2k\pi}{N} \right)},
\label{eq:disc-evalues}
\end{align}
where the distinct eigenvalues are obtained 
by selecting a suitable set of distinct values of~$k$, 
such as for example  
$k=[(1-N)/2],\ldots,[(N-1)/2]$, where $[x]$ stands for the largest integer 
that is less than or equal to~$x$. 

To compare with the continuum Dirac operator, 
we note that the eigenvalues \eqref{eq:disc-evalues} 
of $iM$ have the large $N$ expansion 
\begin{align}
iM: \quad \mu_k &= 1-e^{-i\left( \frac{\theta +2\pi k}{N} \right)} 
 =i\left(\frac{\theta +2\pi k}{N}\right)
+O\left(\left(\frac{\theta +2\pi k}{N}\right)^2\right) , 
\label{eq:disc-evalues-expansion}
\end{align}
while the eigenvalues of 
$i\slashed{\D}$ are
\begin{align}
i\slashed{\D}: \quad \mu_k 
= i\left(\frac{\theta+2\pi k}{l}\right) , 
\quad k \in\Z. 
\label{eq:cont-evalues}
\end{align}
The expressions \eqref{eq:disc-evalues-expansion} and \eqref{eq:cont-evalues}
coincide to $O(N^{-2})$ if the circle has length $l=N$ and $k$ is held
fixed. The matrix $iM$ hence approximates the operator $i\slashed \D$ in
the sense that the eigenvalues of small modulus 
coincide in the limit of a large circle. 

Note from \eqref{eq:disc-evalues-expansion} that the eigenvalues of
$M$ are complex but the imaginary part is subdominant as $N\to\infty$ with
fixed~$k$.  This raises the question as to whether it is fruitful to
think of $M$ as a cut-off version of~$\slashed \D$. This question will be revisited in the conclusion of this part.

\subsection{Mass term}
\label{masssection}

In this subsection it is shown that inclusion of a mass term in a `na\"{i}ve' way breaks triangulation independence.
We start now with a massive continuum fermionic action
\begin{align}
S = \int_0^l \dd t\; \overline{\psi}(t) \left(  \slashed{\D} - m \right) \psi(t),
\end{align}
and follow the usual discretisation procedure using \eqref{derivative} and~\eqref{integral}. The state sum is given by
\begin{align}
{\mathbb I}_{\mathrm{S}^1} =  \int \left( \prod_{j=0}^{N-1} d\psi_j d\overline{\psi}_j \right) e^{ \sum_{j=0}^{N-1} \overline{\psi}_j  \left(  (1-im\Delta t) \psi_j -  Q_{j,j+1} \psi_{j+1} \right) },
\end{align}
where $\Delta t = l/N$, $\psi_{N} = \psi_0$, and the mass term has been discretised according to 
\begin{align}
\int_0^l m\overline{\psi} \psi\;\dd t \rightarrow m\Delta t \sum_{j=0}^{N-1}  \overline{\psi}_j \psi_j.  
\label{mass}
\end{align}
We can evaluate ${\mathbb I}_{\mathrm{S}^1}$ in a similar way as before to obtain
\begin{align}
{\mathbb I}_{\mathrm{S}^1} = \det \left((1-im\Delta t)^N - Q \right),
\end{align}
which is clearly not triangulation independent. The triangulation dependence appears to have crept in with the introduction of a fixed mass scale~$m$. With $m$ fixed, the limit $N \to \infty$ does however yield a well-defined answer,  
\begin{align}
\lim_{N \to \infty}  \det \left( (1-im\Delta t)^N - Q \right)  = \det (e^{- i m l} - Q), 
\label{limit}
\end{align}
where $Q$ is the holonomy around the circle. 

It is possible to include a mass term in the state sum model while maintaining triangulation independence by using the $\U(1)$ matrices $Q'_{i,i+1} = e^{-iA'\Delta t} \; \forall i$ in the state sum model, with $A' = A-m$. In this approach, the mass parameter $m$ is treated as a $\U(1)$ gauge field. Then the partition function of the state sum model is exactly equal to the corresponding continuum functional integral with a mass term, which comes out as
\begin{align}
{\mathbb F}^{A'}_{S^1} = \det( 1-e^{ 2\pi i m}Q).
\end{align}
This construction can be straightforwardly generalised to the $\U(n)$ case.

\subsection{Fermion Doubling}

Fermion doubling is a fairly generic feature of fermionic lattice quantum field theory. The essence of the problem is this: when one na\"{\i}vely calculates the lattice $n$-point functions for a single fermion species in $3 + 1$ dimensions, one finds that in the limit in which the lattice structure is removed, the continuum $n$-point functions are not recovered. Instead one finds a number of spurious states, such that ones ends up with the $n$-point function for $2^4=16$ species of fermions. Nielsen and Ninomiya formulated a theorem \cite{fermion doubling 1, fermion doubling 2, fermion doubling 3} concerning the general conditions under which one can expect this phenomenon to occur.

\begin{theorem} \emph{(Nielsen-Ninomiya)} It is impossible to have a lattice action in $3+1$ dimensions that has the Dirac action as its continuum limit and that simultaneously satisfies all of the following properties:

\begin{enumerate}
	\item Invariance under the global symmetries of the continuum theory (i.e. chiral symmetry).
	\item Local, in the sense that the Fourier transform $i \tilde{M}$ of the fermion matrix $iM$ is a regular function in the momentum space Brillouin zone.
	\item Lattice translation invariance.
	\item Realness.
	\item Bilinearity in the fermion fields.
	\item Free from fermion doubling.
\end{enumerate}
\end{theorem}

I do not know if/how the theorem extends to an odd number of spacetime dimensions. In that case, there is no notion of chirality since the relevant spin group does not decompose into left and right handed parts. However, the proof of the no-go theorem seems to use chirality in an essential way. Therefore I am unaware if there is a version of the theorem that applies in the current context.

Nonetheless the state sum model and the continuum functional integral for the fermionic theory in this chapter are exactly equal. If there is a version of the Nielsen-Ninomiya theorem that applies in an odd number of spacetime dimensions, it is interesting to speculate about how fermion doubling might be avoided under those circumstances.

One possible answer lies in the fact that the discrete action of the state sum model is not real. Note here that complex conjugation is defined for complex Grassmann variables such that the order of the factors is switched: $ (ab)^{\dagger} = b^{\dagger}  a^{\dagger} $. According to this convention, the continuum action $S =  \int dt \overline{\psi}(t) \slashed{D}  \psi(t)$ is real. Crucial to the proof of this is the fact that the derivative $\frac{\dd}{\dd t}$ is an antihermitian operator. However, in the state sum model the discrete version of the derivative operator is

\begin{align}
\frac{\hat \dd}{\dd t} = \frac{T(\Delta t) - 1}{\Delta t},
\end{align}
where the translation operator is defined by $T(\Delta t) \psi_i = \psi_{i+1}$, and transforms under hermitian conjugation as $T^{\dagger}(\Delta t) = T(-\Delta t)$. Therefore $\frac{\hat \dd}{\dd t}$ is clearly not an antihermitian operator. As a result, the discrete action $ \hat{S} = \overline{\psi} iM \psi$ is not real because $iM$ is not hermitian.

This is a very curious fact. Typically in quantum field theory a complex-valued action could lead to major consistency issues, such as complex-valued energies and non-unitary time evolution. However, in a diffeomorphism invariant theory, this may be less problematic. It is an interesting question as to what extent fermion doubling might be avoided in the diffeomorphism invariant setting by the use of complex-valued actions. I leave this question for future investigation.

\chapter{A topological state sum model for a scalar field on the circle} \label{sec: scalar}
In this chapter, the analogue of the one-dimensional state sum model for fermions is developed for the scalar field. In section \ref{sec: scalar state sum}, the definition of the model is developed on the interval and circle. The resulting partition functions are triangulation independent and depend only on the holonomy of the gauge field. In section \ref{sec: scalar interpretation}, it is shown that the state sum model has a discrete action which gives the action for a scalar field minimally coupled to a background gauge field in the continuum limit.

In section \ref{sec: scalar functional}, the partition function of the continuum theory on the circle coupled to an $\OO(n)$ gauge field is evaluated using the zeta-function regularisation detailed in \ref{zetasection}. This gives a result that is identical to the state sum model. 

Introducing a mass term in a `na\"{i}ve' way breaks the triangulation independence of the model. However, a mass term can be incorporated while maintaining the triangulation independence of the partition function on the circle if the gauge group is taken to be the group of strictly positive real numbers under multiplication, with the mass parameter being identified with the gauge field. In this case, the partition function of the state sum model is exactly equal to that of the harmonic oscillator.

\section{The state sum model}\label{sec: scalar state sum}

\begin{figure}[h!]  
\begin{center}
\includegraphics[scale=0.7]{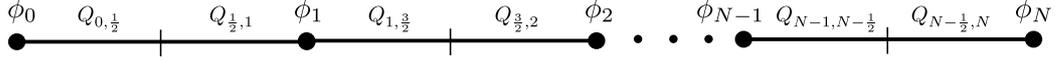}
\end{center} 
\caption{The state sum model for a scalar field on the interval}\label{interval figure}
\end{figure}

Start with an oriented interval $[0,l]$ of length $l$, triangulated with $N+1$ vertices. The vertices are decorated with variables $\phi_i, i=0\ldots N$, each of which is a vector in $\R^n$. The edge connecting the $i$-th and $(i+1)$-th vertices is further subdivided into two segments by a vertex at its centre labelled by $i+\frac{1}{2}$. Each segment with initial vertex $\alpha$ and final vertex $\beta$ is decorated with a real $n \times n$ matrix $Q_{\alpha,\beta}$. Indeed we will use the more general notation that $Q_{\alpha,\beta}$ is equal to the product of the matrices connecting vertices $\alpha$ and $\beta$, which need not be adjacent, in the order determined by the orientation. These matrices satisfy $Q_{\alpha,\beta} = Q_{\beta,\alpha}^{-1}$. The length of each edge is $\Delta t = \frac{l}{N}$. For now we assume that the matrices $Q_{\alpha,\beta}$ are orthogonal. The situation is depicted in figure \ref{interval figure}.

For $N=1$, i.e. a single edge, the state sum model is

\begin{align}
\Z_1 = \left( \frac{1}{2\pi \Delta t} \right)^{\frac{n}{2}} e^{ -\frac{1}{2\Delta t} (Q_{\frac{1}{2},0} \phi_0 - Q_{\frac{1}{2},1} \phi_1  )^2},
\end{align}
with $\Delta t = l$. Gluing two edges together is carried out by multiplying their respective partition functions together and integrating over the variable associated to the interior vertex,

\begin{align}
\Z_2 = \left( \frac{1}{2\pi \Delta t} \right)^{n} \int \dd \phi_1 \; e^{ -\frac{1}{2\Delta t} (Q_{\frac{1}{2},0} \phi_0 -  Q_{\frac{1}{2},1} \phi_1 )^2} e^{ -\frac{1}{2\Delta t} (Q_{\frac{3}{2},1}\phi_1 -  Q_{\frac{3}{2},2} \phi_2)^2}, \label{z2}
\end{align} 
where the integral is the Lebesgue integral over $\R^n$, and now $N=2$, $\Delta t = \frac{l}{2}$. This can be evaluated using the following lemma,

\begin{align}
 \int \dd u \; e^{ -\frac{1}{2a} (x  - M_1 u)^2 } e^{ -\frac{1}{2b} (M_2 u  -  y)^2} = \left(\frac{2\pi ab}{a+b}\right)^{\frac{n}{2}} e^{ -\frac{1}{2(a+b)} (M_1^T x -  M_2^T y)^2},
\end{align}
where $x,y,u \in \R^n$, $a,b>0$ are real numbers , $M_1, M_2$ are real, orthogonal $n \times n$ matrices and the superscript $T$ denotes the matrix transpose. This can be proved by expanding the brackets and using gaussian integration.

Applying this to \eqref{z2} results in

\begin{align}
\Z_2 &= \left( \frac{1}{4 \pi \Delta t} \right)^{\frac{n}{2}} e^{ -\frac{1}{4\Delta t} (Q_{1,0}\phi_0  - Q_{1,2} \phi_2)^2}. \label{Pachner}
\end{align}
Gluing $N$ edges together in this way yields the definition of the state sum model on an interval triangulated with $N+1$ vertices,

\begin{align}
\Z_{[0,l]} &= \left( \frac{1}{2\pi \Delta t} \right)^{\frac{Nn}{2}} \int \prod_{i=1}^{N-1} \dd \phi_i \; e^{ -\frac{1}{2\Delta t} \sum_{i=0}^{N-1} (Q_{i+\frac{1}{2},i}\phi_{i} - Q_{i+\frac{1}{2},i+1} \phi_{i+1})^2} \label{interval0} \\
&= \left( \frac{1}{2\pi N \Delta t} \right)^{\frac{n}{2}} e^{ -\frac{1}{2 N \Delta t} (Q_{m,0} \phi_0  -  Q_{m,N} \phi_{N})^2}  \\
&=  \left( \frac{1}{2\pi l} \right)^{\frac{n}{2}} e^{ -\frac{1}{2l} ( Q_{m,0}\phi_0  -  Q_{m,N} \phi_{N})^2}. \label{scalar interval}
\end{align}
Here $m$ denotes the vertex at the midpoint of the interval.

Assuming $Q_{m,0} \neq Q_{m,N}$ , the state sum model on the circle is obtained by identifying $\phi_{N}=\phi_0$ in \eqref{scalar interval} and integrating to give

\begin{align}
\Z_{S^1} &= \frac{1}{\left|\det(Q_{m,0}-Q_{m,N})\right|}. \label{scalar circle}
\end{align}
In contrast to \eqref{scalar interval}, this result still holds in the important case where $Q_{i,i+\frac{1}{2}} = Q_{i+\frac{1}{2},i+1} =q I \; \forall i $ with $q \in \R \setminus \{ 0,1\}$, and $I$ is the $n \times n$ identity matrix.

The partition functions \eqref{scalar interval} and \eqref{scalar circle} have the property of triangulation independence. That is, they are independent of $N$, the number of sides of the polygon. 

In the case where $Q_{0,m}=Q_{m,N} = e^{ \frac{1}{2} \theta X}$ is an element of $\SO(2)$, with $X=\left( \begin{array}{cc}
0 & -1\\
1 & 0 \\
\end{array} \right)$, $\theta \neq 0$, the partition function \eqref{scalar circle} is given by

\begin{align}
\Z_{S^1} = \frac{1}{4 \sin^2  \frac{\theta}{2}} \label{so(2) result}.
\end{align}
More generally, if $Q_{m,0}$, $Q_{m,N}$ are orthogonal matrices with $Q_{0,m}=Q_{m,N}$, they may be simultaneously diagonalised to the following canonical form

\begin{align}
\left( \begin{array}{cccccc}
 R_1 & & & & & \\
 & \ddots & & & & \\
 & & R_k & & &\\
 & & & \pm 1 & &  \\
 & & & & \ddots & \\
 & & & & & \pm 1 \end{array} \right), 
\end{align}
where the $R_i$, $i=1..k$ are independent $2 \times 2$ rotation blocks. It is clear that under certain circumstances the denominator in \eqref{scalar circle} can be zero, in which case the model is not defined. This occurs if $Q_{0,m}=Q_{m,N}$ is odd-dimensional. Thus this scenario is excluded from consideration. Then the partition function \eqref{scalar circle} decomposes as the product of a number of $\SO(2)$ theories.

In the case where $Q_{0,m}= Q_{m,N} = e^{ -\frac{1}{2} \psi I}$ with $\psi \in \R \setminus \{0\}$, the partition function \eqref{scalar circle} is

\begin{align}
\Z_{S^1} = \left( \frac{1}{\left|2 \sinh \frac{\psi}{2}\right|} \right)^n \label{SHO circle result}.
\end{align}

The state sum model presented here can be straightforwardly generalised to the case where the $\phi_i$ are complex vectors. In this case, where it occurs the orthogonality requirement for the matrices $Q_{\alpha,\beta}$ is replaced by unitarity, and the partition functions \eqref{scalar interval}, \eqref{scalar circle} are the same but for twice as many real degrees of freedom. Due to the fact that any unitary matrix may be diagonalised with complex numbers of unit modulus along the diagonal, the state sum model on the 
circle is defined for unitary matrices $Q_{m,0}$, $Q_{m,N}$ of any dimension.

\subsection{Interpretation of the state sum model}\label{sec: scalar interpretation}

The state sum model of this section may be interpreted as the partition function of a minimally coupled, real scalar field theory.

The $\phi_i$'s are interpreted as a discrete sampling of a continuous real scalar field $\phi(t)$, with $t\in [0,l)$, so that

\begin{align}
\phi_j = \phi\left(j\Delta t \right).
\end{align}
The interval connecting vertices $\alpha$ and $\beta$ is assigned the matrix $Q_{\alpha,\beta}$, which is interpreted as the parallel transporter for the $\phi$ field along that interval. Then $Q_{\alpha,\beta} = e^{\int_{\alpha \Delta t}^{\beta \Delta t} \dd t  A(t)} $, with $A(t)$ the gauge field on the $1$-manifold. Up to a minus sign, the argument of the exponent in \eqref{interval0} may be seen to be a lattice discretisation of the usual action for a scalar field,

\begin{align}
\hat{S} &=  \frac{1}{2\Delta t} \sum_{i=0}^{N-1} \left(Q_{i+ \frac{1}{2},i }\phi_{i} - Q_{i+\frac{1}{2},i+1} \phi_{i+1}\right)^2  \nonumber \\
&=  \frac{\Delta t }{2} \sum_{i=0}^{N-1} \left(\frac{Q_{i+ \frac{1}{2},i }\phi_{i} - Q_{i+\frac{1}{2},i+1} \phi_{i+1} }{\Delta t}\right)^2 .\label{discrete action 2}
\end{align}
The limit $N \rightarrow \infty$, or equivalently $\Delta t \rightarrow 0$, while keeping $l$ constant can be evaluated,

\begin{align}
\lim_{\Delta t \rightarrow 0} \hat{S} &= \lim_{\Delta t \rightarrow 0} \frac{\Delta t }{2} \sum_{i=0}^{N-1} \left(\frac{Q_{i+ \frac{1}{2},i }\phi_{i} - Q_{i+\frac{1}{2},i+1} \phi_{i+1} }{{\Delta t}}\right)^2  \nonumber  \\
&= \frac{1}{2} \int_0^l \dd t \; (\D\phi)^2, \label{continuum limit}
\end{align}
where $\D = \frac{\dd}{\dd t}  + A$ is the covariant derivative. This is just the usual continuum action for a scalar field. Thus the partition function \eqref{interval0} may be interpreted as that of a real, minimally coupled scalar field.

It is possible to add a mass term to the action \eqref{discrete action 2},

\begin{align}
\hat{S}_m = \frac{1}{2} m^2 \Delta t \sum_{i=0}^{N-1} \phi_i^2 . \label{naive mass term}
\end{align}
However, the resulting partition functions on the interval and circle are no longer triangulation independent.

If the gauge group is taken to be isomorphic to the abelian group of strictly positive real numbers under multiplication so that $A(t) \sim I$, then

\begin{align}
\frac{1}{2} \int_0^l \dd t \; (\D\phi)^2 = \frac{1}{2} \int_0^l \dd t \left[ \left( \frac{\dd \phi}{\dd t} \right)^2 + \alpha^2 \phi^2 \right]. \label{massive scalar}
\end{align}
Here the gauge freedom has been used to transform the gauge field $A(t)$ so that it is everywhere equal to a constant $\alpha I$. Identifying $\phi(t)$ with the $n$-dimensional position vector $x(t)$ and $\alpha$ with the spring constant reveals that this is precisely the Euclidean action for the simple harmonic oscillator. Alternatively the action may be viewed as that of a massive scalar field upon identifying $\alpha$ with the mass parameter. Thus it is possible to introduce a mass term into the state sum model if the mass parameter is treated as an element of the Lie algebra $\R$. The corresponding matrices $Q_{\alpha,\beta}$ furnish an $n$-dimensional representation of the abelian group of strictly positive numbers under multiplication, as is the case in \eqref{SHO circle result}. In this way it is possible to include a mass term in the state sum model on the circle while maintaining triangulation independence.

\section{Comparison with functional integral}\label{sec: scalar functional}
In this section, we compute the zeta function regularised partition function of the continuum theory for the gauge group $\SO(2)$ using the machinery of subsection \ref{zetasection}, and show that it is equal to the result from the state sum model. 

For a minimally coupled scalar field theory, we may define

\begin{align}
\Z &= \int \mathcal{D} \phi \; e^{- \frac{1}{2} \int \phi \cdot L \phi } \nonumber \\
&:= \frac{1}{{\sqrt{\det L}}}, \label{continuum partition function}
\end{align}
where the square root is the positive square root and the determinant is defined by zeta function regularisation. The differential operator $L$ gives the classical action in the exponent. In the case of a real scalar field minimally coupled to an $\SO(2)$ gauge field on the circle, we have

\begin{align}
S &= \frac{1}{2} \int_0^l \dd t \; (\D\phi)^2 \nonumber \\
&=  \frac{1}{2} \int_0^l \dd t \; \phi \cdot \left( -\frac{\dd^2}{\dd t^2} -2A \frac{\dd}{\dd t} - A^2 \right) \phi. \label{action1}
\end{align}
Therefore in this case we have $L= -\frac{\dd^2}{\dd t^2} -2A \frac{\dd}{\dd t} - A^2$. 
 
In order to compute $\det L$, we first need to know the eigenvalues of $L$. The gauge freedom may be used to transform the gauge field $A(t)$ so that it is everywhere equal to a constant $\alpha$, $A = \alpha \left( \begin{array}{cc}
 0 & -1 \\
 1 & 0 \\
 \end{array} \right)$, with $\alpha= \frac{2\pi a}{l}$ and $a \in [0,1)$. Then the eigenvalue equation is

\begin{align}
\left( \begin{array}{c}
 -\frac{\dd^2 v_1}{\dd t^2} +2\alpha \frac{\dd v_2}{\dd t} + \alpha^2 v_1  \\
 -\frac{\dd^2 v_2}{\dd t^2} -2\alpha \frac{\dd v_1}{\dd t} + \alpha^2 v_2 \\
 \end{array} \right)
= 
\lambda \left( \begin{array}{c}
  v_1  \\
  v_2 \\
 \end{array} \right),
\end{align}
and we impose periodic boundary conditions, $v_{1,2} (t+l) =v_{1,2} (t) $. This has linearly independent solutions $\left( \begin{array}{c}
  \cos \frac{2\pi kt}{l}  \\
  \sin \frac{2\pi kt}{l} \\
 \end{array} \right)$ and $\left( \begin{array}{c}
  \sin \frac{2\pi kt}{l}  \\
  \cos \frac{2\pi kt}{l} \\
 \end{array} \right)$, $k \in \Z$. The eigenvalues are given by
 
 \begin{align}
 \lambda_{k_{\pm}} = \left(\frac{2\pi}{l}(k \pm a)\right)^2, \;\; k \in \Z.
 \end{align}
The determinant in \eqref{continuum partition function} is defined by 

\begin{align}
\det L = e^{- \zeta_L'(0) },
\end{align}
with 

\begin{align}
\zeta_L(s) &=  \left(\frac{2\pi}{l}\right)^{-2s} \sum_{k \in \Z} \left(  \frac{1}{(k+a)^{2s}} + \frac{1}{(k-a)^{2s}} \right) \nonumber \\
&= \left(\frac{2\pi}{l}\right)^{-2s} \sum_{k \in \Z} \frac{2}{(k+a)^{2s}}. \label{zeta}
\end{align}
This may be re-written in terms of the Hurwitz zeta function,

\begin{align}
\zeta_L(s) =  \left(\frac{2\pi}{l}\right)^{-2s} \Big( 2\zeta_H(2s, a) + 2\zeta_H(2s, 1-a)\Big),
\end{align}
where $\zeta_H$ is defined by

\begin{align}
\zeta_H(s, q) = \sum_{k=0}^{\infty} \frac{1}{(k+q)^s}.
\end{align}
The Hurwitz zeta function can be analytically continued to remove the pole at $s=0$. Using \eqref{zeta stuff}, we have

\begin{align}
\zeta'_L(0) = 4 \ln \Gamma(a) + 4 \ln \Gamma(1-a) -4 \ln 2\pi.
\end{align}
Therefore,

\begin{align}
e^{- \zeta'_L(0)} &= e^{-4 \ln (\Gamma(a) \Gamma(1-a)) + 4\ln 2\pi} \nonumber\\
&= \left( \frac{2\pi}{\Gamma(a) \Gamma(1-a)} \right)^4 \nonumber \\
&= (4 \sin^2 \pi a)^2.
\end{align}
The partition function \eqref{continuum partition function} is

\begin{align}
\Z = \frac{1}{4 \sin^2 \pi a}.
\end{align}
The holonomy is given by $Q=e^{2\pi a}$, and identifying $\theta = 2\pi a$ gives precisely the result \eqref{so(2) result} from the state sum model. Thus the zeta function regularised functional integral is equivalent to the state sum model. This result generalises immediately to the $\OO(n)$ case by diagonalising the connection with a gauge transformation, whereupon the the functional integral is the product of a number of $\SO(2)$ functional integrals. 

The action for the Euclidean harmonic oscillator in $n$ spatial dimensions is

\begin{align}
S = \frac{1}{2} \int_0^l \dd t \; x(t) \left( -\frac{\dd^2}{\dd t^2} + \omega^2 \right) x(t). \label{SHO}
\end{align}
The corresponding partition function on the circle is calculated in \cite{HO},

\begin{align}
{\mathbb Z}_{S^1} = \left( \frac{1}{2\sinh\frac{\omega l}{2}}\right)^n,
\end{align}
where $\omega > 0$ is understood to be the positive square root of $\omega^2$. This result is identical to the state sum model \eqref{SHO circle result} after identifying $\psi=\omega l$.

\section{Discussion} \label{discsection}
In this part, we have constructed one-dimensional state sum models for fermionic and scalar fields on the interval and the circle. The resulting partition functions are simple functions of the holonomy that are triangulation independent. We have carried out an exact calculation of the partition functions in the continuum using zeta function methods, and have obtained results identical to the state sum models. A precise comparison has been made between the discrete state sum model and the functional integral in the partition function of the continuum theory.

A curious feature of the partition functions \eqref{interval} and \eqref{circle} for the fermionic field is that they do not depend on the length of the circle, whereas the eigenvalues of both the continuum Dirac operator $\slashed{\D}$ and the discrete Dirac operator $M$ clearly do.

Some more insight into this can be gained by examining an operator cut-off regularisation of the Dirac functional integral. Such a regularisation depends on a cut-off scale $c$. It replaces $i\D$ with $f(i\D)$, where the function $f(z)$ of a complex variable is the identity function $f(z)=z$ for $|z|\ll c$, but $f(z)=1$ for $|z|\gg c$. Thus it effectively removes the eigenvalues of $\D$ with magnitude above the cut-off $c$. An example of such a cut-off regularisation is the Schwinger proper time regularisation. 

As $c\to\infty$, the regularised determinant diverges. A calculation for the Schwinger proper time regularisation shows that the leading asymptotic term for $\log\det |\slashed{\D}|$ is $(l/\pi)\,c\log c$. This can be confirmed in a simple way for special cases, e.g.\ $a=1/2$, using a sharp cut-off and Stirling's formula for the asymptotic expansion of the factorial. The cut-off regularisation only agrees with the zeta function method once the leading asymptotic terms are removed \cite{baer-schopka}. It is worth noting that these leading divergent terms are proportional to $l$, thus explaining the role played by the length of the circle in evaluating this determinant. Thus to get a partition function that converges as $c\to\infty$, one can multiply the cut-off regularised determinant with a `cosmological term' 
$e^{\Lambda(c) l}$, choosing a suitable function $\Lambda(c)$ to renormalise the 
coefficient of $l$ as $c\to\infty$. This is the same as adding a term $\Lambda(c) l$ to the exponent in \eqref{Dirac generating functional} when using an operator cut-off regularisation. The functional integral formula for this regularisation is then  
\begin{align}
\int \mathcal{D} \psi \mathcal{D} \overline{\psi}  \, e^{i\int_0^{l} \dd t \, (\overline{\psi}(t) \slashed{\D} \psi(t)-i\Lambda)}.
\end{align} 
Similar divergences do not occur with cut-off regularisations of the eta invariant in the phase term \cite{bismut-freed}, so there are no additional parameters to renormalise for the phase of the partition function.

Our conclusion from this discussion is that the state sum model is a more subtle regularisation of the determinant of the Dirac operator than a mere operator cut-off. 

There are some striking differences between the state sum models for the fermionic and scalar field. The fermionic state sum model is well defined for any gauge group, whereas the partition function for the real scalar field on the circle \eqref{scalar circle} is not well defined for the gauge group $\SO(2n+1)$. In the case of a complex scalar field however, the model is defined for any unitary group.

Furthermore, in the fermionic case, the partition function for the circle \eqref{circle} may be sensitive to the orientation for certain gauge groups. By contrast, the partition function \eqref{scalar circle} never detects an orientation.

Introduction of a mass term in a na\"{\i}ve way breaks the triangulation independence of the models. However, one interesting result from this work is that a mass term can still be accommodated in a triangulation independent way if the mass parameter is treated as a gauge field for the appropriate group.

In the standard treatment of the path integral for the harmonic oscillator, see e.g. \cite{Feynman}, the partition function is calculated as the limit of a discrete model that is not triangulation independent. The state sum model \eqref{scalar circle} has the virtue that it is triangulation independent and exactly equal to the partition function of the harmonic oscillator once the appropriate gauge group has been chosen.

For one-dimensional manifolds, the choice between Euclidean and Lorentzian signatures is a matter of convention, and the analogous results for the other metric signature can be obtained by analytically continuing the results in this part.

\part{Gauge gravity and quantisation}

In this part we consider the formulation of gravity as a gauge theory, and its quantisation in $2+1$ and $3+1$ dimensions. In $2+1$ dimensions, the classical first order theory of gravity is equivalent to a Chern-Simons theory with gauge group $\ISO(2,1)$ for zero cosmological constant and $\SO(3,1)/\SO(2,2)$ for positive/negative cosmological constant. In \cite{Witten, Witten 2}, Witten quantised $(2+1)$-dimensional gravity exploiting the fact that it is equivalent to a Chern-Simons gauge theory.  

There have been various attempts at realising gravity as a gauge theory in $3+1$ dimensions. Sciama and Kibble discovered that gravity in $3+1$ dimensions can be written in a form that exhibits a local $\SO(3,1)$ gauge symmetry using the frame field formalism \cite{Kibble, Sciama}. In the Macdowell-Mansouri formulation \cite{MM}, the action has an $\SO(4,1)/\SO(3,2)$ symmetry that is explicitly broken to obtain the Sciama-Kibble first order form of the gravitational action plus an instanton term. In this approach, the frame field and spin connection are assimilated together as components of the $\SO(4,1)/\SO(3,2)$ connection. Stelle and West \cite{SW} expanded on this work by including an additional scalar field to carry out the symmetry breaking, thus restoring the overall $\SO(4,1)/\SO(3,2)$ symmetry when all fields are considered. This action was improved by Pagels \cite{Pagels}, giving a simpler action principle and removing the gravitational instanton term. Pagels also described how to couple the theory to scalar, fermion and Yang-Mills fields. A Poincar\'e group analogue of Pagels' work was developed by Grignani and Nardelli \cite{GN}, and the coupling to scalar and Yang-Mills fields improved upon by Ha \cite{Ha}.

In chapters \ref{chap: Gauge gravity action} and \ref{chapter: Coupling to matter} these developments are reviewed, but using a simplified matrix formalism for the $\ISO(n)$ case that is similar to Pagels' original $\SO(n+1)$ theory. Initially, Euclidean signature is chosen for simplicity, so that in $n$ spacetime dimensions the gauge group is the Euclidean group $\ISO(n)$ for zero cosmological constant and the special orthogonal group $\SO(n+1)$ for non-zero cosmological constant. However, the generalisation to other signatures is immediate. In the actions constructed here, the spin connection and the frame field are packaged as parts of an $\ISO(n)/\SO(n+1)$-valued connection form, and there is a scalar field that is charged under the gauge group that effects a reduction in symmetry to $\SO(n)$, giving the usual first-order form of the Einstein-Hilbert action. The symmetry breaking is the same mechanism as in the broken phase of a spontaneously-broken gauge theory. The scalar field can be thought of as a Higgs field that is constrained to lie in its vacuum manifold; the model does not have an unbroken phase. The coupling to scalar and Yang-Mills fields that was proposed by Ha in \cite{Ha} for the $\SO(n+1)$ case is generalised to $\ISO (n)$. The fermionic $\ISO(n)$ coupling that is proposed here differs from that of Grignani and Nardelli \cite{GN} because the spinors are placed in a representation of $\ISpin(n) = \Spin(n) \rtimes \R^n$ in which the translations are represented trivially. This leads to a simpler action principle. We also show that it is not possible to obtain chiral fermions in an even number of spacetime dimensions within the $\SO(n+1)$ formalism.

In chapter \ref{chapter: Quantisation}, a sum over histories quantisation of the $\ISO(3)/\ISO(2,1)$ theory in $2+1$ dimensions is carried out. This discrete model is novel and depends on a certain geometric structure (a set of loops) for which we do not yet have a general definition that is valid for any topology of space-time. However it is possible to specify this structure, and hence the model precisely, for collapsible triangulations of the three-sphere. It is shown that for these triangulations it reduces to the Ponzano-Regge model, with the structure specifying exactly the gauge-fixing required for the definition of the Ponzano-Regge model. The model is then extended to Lorentzian signature, using as gauge group the Poincar\'e group in three dimensions.

Finally, in chapter \ref{chapter: Hamiltonian} the Hamiltonian analysis of the $(3+1)$-dimensional gauge gravity theory is begun. The `na\"{i}ve' Hamiltonian takes a simple form in the gauge gravity formalism. The Hamiltonian analysis is only at a preliminary stage, and its completion is a challenge left for future work.

\chapter{Gauge gravity action}\label{chap: Gauge gravity action}
In this chapter, the gauge action for gravity in $n$ spacetime dimensions is reviewed. For simplicity, the $\SO(n+1)$ and $\ISO(n)$ theories will be presented. The $\SO(n+1)$ theory is given in \cite{Pagels}, and the $\ISO(n)$ theory in \cite{GN}. However, we use a simple matrix formalism for the $\ISO(n)$ case. The $\SO(n+1)$ theory naturally leads to the first order form of the Einstein-Hilbert action with a negative cosmological constant, while the $\ISO(n)$ theory has zero cosmological constant. The signature of the groups may be altered to accommodate different spacetime signatures and a different sign of the cosmological constant, i.e. $\SO(n,1)$ or $\SO(n-1,2)$ as opposed to $\SO(n+1)$.

\section{$\SO(n+1)$ action - $\Lambda \neq 0$}\label{sec:so(n) gauge gravity}

The Lie algebra for the group $\SO(n+1)$ can be represented as real, antisymmetric $(n+1) \times (n+1)$ matrices. The $\SO(n+1)$ connection is denoted $A^{BC}$, and the curvature is ${F}^{BC}=\dd A^{BC} +A^{BD}\wedge A_D^{\phantom{D}C}$, with upper case indices $A,B \ldots=0,\ldots,n$ and repeated indices contracted.
Indices are raised and lowered with the Euclidean metric $\delta_{AB}$. We introduce a multiplet of scalar fields $\phi^A$ taking values in a sphere in $\R^{n+1}$ with constant radius $c>0$, so that $\phi^A\phi_A=c^2$. The covariant derivative is
\begin{align}
\D\phi^B=\dd \phi^B+{A^B}_C \phi^C. \label{covariant}
\end{align}
Pagels' action is \cite{Pagels}
\begin{align}\label{pagelsaction}
S = \int & (\D \phi)^A\wedge (\D \phi)^B\wedge\ldots (\D \phi)^W 
\wedge  {F}^{XY} 
\epsilon_{AB\ldots WXYZ} \phi^Z. 
\end{align}
In $n$ spacetime dimensions, there are $(n-2)$ instances of $\D\phi$ in the action. For example, the $(3+1)$-dimensional action is 
 
\begin{align}
S = \int & (\D \phi)^A 
\wedge  (\D \phi)^B 
\wedge {F}^{CD} 
\epsilon_{ABCDE} \phi^E. 
 \label{new action}
\end{align}

The $\phi$ field is a vector in $\R^{n+1}$ and so, by a gauge transformation, it may be rotated so that it points along the final coordinate axis,
\begin{align}\label{physicalgauge}
 \phi^Z \rightarrow \left( \begin{array}{c} 0 \\
 \vdots \\
 0 \\
 c \\
 \end{array} \right).
 \end{align}
This gauge choice is known as `physical gauge'. In this gauge, the fields may be written in terms of $\SO(n)$ tensors in block form, 

\begin{align}\label{physicalgauge2}
 \phi^Z=\begin{pmatrix}0\\c\end{pmatrix}
 \end{align}
with $0\in\R^n$, $c\in\R$, and

\begin{align}\label{connection}
A^{BC} =\begin{pmatrix}\omega^{bc}&e^b\\-e^c&0\end{pmatrix}. 
\end{align}
Here $\omega^{bc}$ is an $n\times n$ matrix of one-forms and $e^b$ is an $n$-dimensional vector of one-forms. Capital indices $A,B \ldots=0\ldots n$ are in the fundamental representation of $\SO(n+1)$, and the corresponding lower case indices $a,b \ldots =0\ldots (n-1)$ are in the fundamental representation of the $\SO(n)$ subgroup.  Thus $A^{bc}=\omega^{bc}$, $A^{bn}=e^b$, $A^{nc}=-e^c$ $A^{nn}=0$. 
 
In the physical gauge, the 1-forms $e^b$ and $\omega^{bc}$ are interpreted as the components of the frame fields and spin connection respectively. We have

\begin{align}
 (\D\phi)^B = \begin{pmatrix}ce^b\\ 0\end{pmatrix}.
 \end{align}
 Defining $R^{ab}$ to be the curvature of the  $\SO(n)$-connection $\omega$, we have
\begin{equation} F^{ab}=R^{ab}- e^a\wedge e^b.\end{equation}
Thus the action \eqref{pagelsaction} is
 \begin{align}\label{gravityaction}
 S = c^{n-1} \int  e^a  \wedge e^b\wedge\ldots \wedge e^w \wedge R^{xy}\, \epsilon_{ab\ldots wxy} - e^a \wedge e^b \wedge\ldots \wedge e^y\, \epsilon_{ab\ldots y} .
 \end{align}
This is the Sciama-Kibble action for gravity with a non-zero cosmological constant. It can be made to take its more familiar form by rescaling the frame field. Defining $\Lambda>0$ by $c^{n-1}\Lambda^{(n-2)/2}=1/G\hbar$ and setting $\tilde e^a=\Lambda^{-1/2}\,e^a$ results in the usual first-order form of the Einstein-Hilbert action with negative cosmological constant,
\begin{align}
 S = \frac1{G\hbar} \int  \tilde e^a  \wedge \tilde e^b\wedge\ldots \wedge \tilde e^w \wedge R^{xy}\, \epsilon_{ab\ldots wxy} -\Lambda \,\tilde e^a \wedge \tilde e^b \wedge\ldots \wedge \tilde e^y\, \epsilon_{ab\ldots y} .
 \end{align}
 The group signature may be altered to $\SO(n,1)$ obtain the same action with a positive cosmological constant.

\section{${\ISO}(n)$  action - $\Lambda = 0$}\label{sec:iso(n) gauge gravity}

In this section, the gauge gravity action for the Euclidean group is constructed by analogy with the action of Pagels. This uses a matrix representation of $\ISO(n)$ that is similar to the defining representation of $\SO(n+1)$.  This results in an action in which the cosmological constant is naturally zero. When written in field components, the action coincides with the action studied in \cite{GN}. First some facts about the representation theory of the Euclidean group are reviewed.

\subsection{Representations of the Euclidean group}
The $n$-dimensional Euclidean group $\ISO(n)$ is the group of rotations and translations of $\R^n$. Its action is given by $x \rightarrow Mx + t$, where $M\in\SO(n)$ is the rotation matrix and $t\in\R^n$ is the translation vector. This defining representation has a non-linear action but can be represented in terms of matrices if the dimension of the matrices is increased by one. The action is
\begin{align}
\left( \begin{array}{c} x \\
c \\
\end{array} \right) &\rightarrow \left( \begin{array}{cc} M & t \\
0 & 1 \\
\end{array} \right) \left( \begin{array}{c} x \\
c \\
\end{array} \right) = \left( \begin{array}{c} Mx+ct \\
c \\
\end{array} \right). \label{vector rep}
\end{align}
This will be called the vector representation, and indices transforming in this representation will be denoted as upper indices. The particular value $c=0$ gives a linear subspace that is a sub-representation in which only the rotations act,
\begin{align}\label{quotientrep}
\begin{pmatrix}x\\0\end{pmatrix}\rightarrow\begin{pmatrix} Mx\\0\end{pmatrix}.
\end{align}

The dual of the vector representation is called the covector representation, and indices transforming in this representation will be denoted as lower indices. 
This may be represented as follows. Let $k=(k_0,\ldots,k_{n-1})\in\R^n$ and $k_n=\Omega\in\R$ be the last coordinate. The action is
\begin{align}
\left( \begin{array}{c} k \\
\Omega \\
\end{array} \right) &\rightarrow \left( \begin{array}{cc} M & 0 \\
-M^{-1}t & 1 \\
\end{array} \right) \left( \begin{array}{c} k \\
\Omega \\
\end{array} \right)= 
 \left( \begin{array}{c} Mk \\
-\left(M^{-1}t\right).k + \Omega \\
\end{array} \right). \label{covector rep}
\end{align}
The invariant contraction of a vector and covector is
\begin{align} x^Ak_A=x\cdot k + c\Omega=
 \left( \begin{array}{cc} Mx+ct & c  \\
\end{array} \right) \left( \begin{array}{c} Mk \\
-\left(M^{-1}t\right).k + \Omega \\
\end{array} \right) = x'^A k'_A .
\end{align}
The primes here denote transformed quantities. 

The invariant bilinear form that can be used to contract two covectors is given by
\begin{align}
\eta^{AB} = \left( \begin{array}{cccc} 1 & & &  \\
&  \ddots & & \\
& & 1& \\
& & & 0 \end{array} \right), \label{upper metric}
\end{align}
since $ k'_A \eta^{AB} l'_{A} = k'\cdot l'=k\cdot l = k_{A} \eta^{AB} l_A $.

The invariant bilinear form that can be used to contract two vectors is given by
\begin{align}
\eta_{AB} = \left( \begin{array}{cccc} 0 & & &  \\
 & \ddots & & \\
& & 0 & \\
& & & 1  \end{array} \right) , \label{lower metric}
\end{align}
since $ x'^A \eta_{AB} y'^A = c^2 = x^A \eta_{AB} y^A $.

Vectors that lie in the sub-representation \eqref{quotientrep} may be contracted using the identity matrix $\delta_{AB}$, since these vectors transform trivially under translations.

Finally, we note that the permutation symbols $\epsilon_{AB\ldots YZ}$ and $\epsilon^{AB \ldots YZ}$ are both invariant, since the transformations $\eqref{vector rep}$ and $\eqref{covector rep}$ both have determinant $1$.

\subsection{The action}
Now an $\ISO(n)$ invariant action will be constructed using these ingredients.  Since the bilinear forms \eqref{upper metric}, \eqref{lower metric} are degenerate, the operations of raising and lowering indices are not invertible and one has to be work out which quantities are naturally vectors or covectors. 

The $\phi$ field is now a multiplet of real scalar fields in the vector representation, with a fixed constant $c\in\R$ as the last component,
$\phi^A = \left( \begin{array}{c} \phi^a \\
c \\
\end{array} \right) $. The $\ISO(n)$ connection is given in block form as

\begin{align}{A^B}_C=\begin{pmatrix}{\omega^b}_c &e^b\\0&0\end{pmatrix}. \label{iso connection}
\end{align}
The covariant derivative \eqref{covariant} is
\begin{equation}\D\phi^B=\begin{pmatrix} \dd\phi^b +{\omega^b}_c\phi^c + ce^b\\0\end{pmatrix}.\end{equation}
Since the last component is zero, this lies in the sub-representation \eqref{quotientrep}, transforming covariantly under rotations and not at all under translations.

Now consider the Euclidean field strength tensor, ${F}=dA + A\wedge A$. In terms of matrix components it is given by
\begin{align}
{F}^B_{\;\;\;C} = \left( \begin{array}{cc} \dd{\omega^b}_c + {\omega^b}_d \wedge {\omega^d}_c & \dd e^b + {\omega^b}_d \wedge e^d  \\
0 & 0  \end{array} \right).
\end{align}
Raising the second index using the metric $\eta^{AB}$ in $\eqref{upper metric}$ gives
\begin{align}
{F}^{BC} = {F}^B_{\;\;\;D} \eta^{DC} = \left( \begin{array}{cc} \dd\omega^{bc} + {\omega^b}_d \wedge \omega^{dc}  & 0  \\
0 & 0  \end{array} \right).
\end{align}
This tensor is also invariant under translations.

Finally an $\ISO(n)$ invariant action can now be constructed using the same formula as Pagels' action, \eqref{pagelsaction},

\begin{align}\label{GNaction}
S = \int & (\D \phi)^A\wedge (\D \phi)^B\wedge\ldots (\D \phi)^W 
\wedge  {F}^{XY} 
\epsilon_{AB\ldots WXYZ} \phi^Z. 
\end{align}
This action can be gauge fixed, as in \eqref{physicalgauge}. It reduces to
  \begin{align} \label{gravityaction2}
 S = c^{n-1} \int e^a  \wedge e^b\wedge\ldots \wedge e^w \wedge R^{xy}\, \epsilon_{ab\ldots wxy}  ,
 \end{align}
which is exactly the Sciama-Kibble action for gravity with zero cosmological constant. The action allows a rescaling of the frame field, which is equivalent to changing the value of $c$. 

A cosmological term may be added to the theory by starting with an additional term in the action proportional to
\begin{align}
 \int \vol',
\end{align}
where the $n$-form $\vol'$ is defined as

\begin{align}
\vol'= (\D\phi)^A \wedge  (\D\phi)^B \wedge\ldots\wedge (\D\phi)^Y  \epsilon_{AB\ldots YZ} \phi^Z. \label{volume form}
\end{align}
There are $n$ instances of $\D\phi$ in this formula. In the physical gauge, it is readily seen that $ \vol' = c^{n+1} \vol=  c^{n+1} e^a \wedge e^b \wedge\ldots\wedge e^y\, \epsilon_{ab\ldots y}$, with $\vol$ the canonical volume form on the spacetime manifold.

\chapter{Coupling to matter}\label{chapter: Coupling to matter}
In this chapter, the coupling of the gauge gravity action to matter is explored. A way of coupling the $\SO(n+1)$ theory to fermion fields has been explored before by Pagels \cite{Pagels}, and subsequently generalised to the $\ISO(n)$ case by Grignani and Nardelli \cite{GN}. The general form of the coupling to scalar and Yang-Mills fields was worked out by Ha in \cite{Ha} for the $\SO(n+1)$ case. We generalise this to the $\ISO(n)$ case. The fermionic $\SO(n+1)$ coupling  presented in this chapter is identical to that of Pagels. However, it is pointed out that it is not possible to obtain chiral fermions in an even number of spacetime dimensions within the $\SO(n+1)$ formalism. The fermionic $\ISO(n)$ coupling presented in this chapter is simpler than that of Grignani and Nardelli insofar as the spinors transform covariantly under rotations and trivially under translations, rather than covariantly under the whole Euclidean group, and the action is written using the simplified matrix formalism of the previous section.  The results here immediately generalise to other signatures, i.e. $\SO(n,1)$, $\SO(n-1,2)$, etc.

\section{Bosons}
In this section it is shown how to couple the gauge gravity action to scalar fields and Yang-Mills fields.

The general action for a real singlet scalar field $\sigma$ in $n$ dimensions is

\begin{align}
S_S =  \frac{1}{2} &\int \vol \; \left[ (\partial^{\mu} \sigma) (\partial^{\nu} \sigma) g_{\mu \nu}  - m^2 \sigma^2 \right], \label{general scalar}
\end{align}
where $g_{\mu \nu}$ is the spacetime metric and $\vol$ is the canonical volume form.

The $\SO(n+1)$ theory may be coupled to a real singlet scalar field $\sigma$ in a gauge invariant way as follows,

\begin{align}
S_S = \frac{1}{2} &\int {\vol'} \; \left[ (\partial^{\mu} \sigma) (\partial^{\nu} \sigma) g_{\mu \nu}'   - m^2 \sigma^2 \right]. \label{scalar coupling}
\end{align}
Here $\vol'$ is defined in \eqref{volume form}, and $g_{\mu \nu}'$ is defined by 

\begin{align}
g_{\mu \nu}' &= (\D_{\mu} \phi)^A (\D_{\nu} \phi)^B \delta_{AB}. \label{gauge metric}
\end{align}
Upon going to the physical gauge,

\begin{align}
 \phi^A \rightarrow \left( \begin{array}{c} 0 \\
 c \\
 \end{array} \right),
 \end{align}
with $c=1$ taken for convenience, and using the fact that the metric and frame field are related by

\begin{align}
g_{\mu \nu} = e_{\mu}^a e^b_{\nu} \delta_{ab},
\end{align}
it is seen that $g_{\mu \nu}' = g_{\mu \nu}$. From \eqref{volume form} we also have that $\vol' = \vol$. Therefore the action \eqref{scalar coupling} is equal to \eqref{general scalar}, the general action for a scalar field coupled to gravity.

This coupling may be straightforwardly generalised to the case where the scalar field is complex or charged under some additional symmetry group. In the latter case, the partial derivatives in \eqref{scalar coupling} are replaced by the appropriate covariant derivatives.

A similar strategy may be used to couple Yang-Mills theory to the gauge gravity action. The general action for a Yang-Mills field in $n$ dimensions is

\begin{align}
S_{YM} = \int \vol \; \tr (F^{\mu \nu} F^{\rho \lambda}) g_{\mu \rho} g_{\nu \lambda}, \label{general Yang-Mills}
\end{align}
where $F^{\mu \nu}$ is the curvature tensor for an appropriate external symmetry group, and $\tr$ is the Killing form on the relevant Lie algebra.
The gauge gravity coupling is given by 

\begin{align}
S_{YM}  =  &\int  \vol' \; \tr(F^{\mu \nu} F^{\rho \lambda}) g_{\mu \rho}' g_{\nu \lambda}',
\end{align}
which immediately reduces to \eqref{general Yang-Mills} in the physical gauge with $c=1$.

Identical constructions works for the gauge group $\ISO(n)$. In that case, equation \eqref{gauge metric} is manifestly $\ISO(n)$ invariant because the vector $(\D \phi)^A$ lies in the sub-representation \eqref{quotientrep}, for which the identity matrix $\delta_{AB}$ is an invariant bilinear form.

\section{Fermions}

In this section, it is shown how to couple the gauge gravity action to fermion fields.

Let the spacetime metric signature be $(p,q)$, with $p$ denoting the number of positive eigenvalues of $g_{\mu \nu}$, and $q$ denoting the number of negative eigenvalues, with $p+q=n$. The massless Dirac action is given by 

\begin{align}
S_F= \frac{\alpha}{2} \int  \;  e^a \wedge & \; e^b \wedge \ldots  e^x \wedge \left[ \overline{\psi} \gamma^y (\dd + \omega)\psi - \left((\dd + \omega) \overline{\psi}\right) \gamma^y  \psi \right]  \epsilon_{ab\ldots xy}. \label{Dirac action}
\end{align}
Here it is necessary to split the action into two terms because the torsion tensor is not assumed to be zero. There are $(n-1)$ instances of the frame field $e$ in this action. The spinor $\psi$ is in a representation of the spin group $\Spin(p,q)$. The integrand in \eqref{Dirac action} can be pure real or imaginary depending on the number of spacetime dimensions, the spacetime metric signature and the choice of spinor inner product. Therefore the constant $\alpha$ is chosen to be proportional to $1$ or $i$ appropriately so that the action is real. Hermitian conjugation in the Grassmann algebra is defined so that the combination $\overline{\psi}\psi$ is real - in other words, one does not pick up an additional minus sign from the Grassmann anti-commutation law in reversing the order of the factors upon conjugation. The gamma matrices generate the real Clifford algebra $\mathrm{Cl}(p,q)$,

\begin{align}
\{ \gamma^a, \gamma^b \} = 2\eta^{ab},
\end{align}
where  $\eta^{ab} = \diag (\overbrace{1,1, \ldots 1}^\text{p times}, \overbrace{ -1, -1, \ldots -1}^\text{q times}) $. The action of the spin connection $\omega$ on spinors is given by $\omega \psi = \omega_{ab} S^{ab} \psi$, with $\omega_{ab}$ real parameters and $S^{ab} = \frac{1}{4} [\gamma^a, \gamma^b]$ the generators of $\Spin(p,q)$. The spinor inner product $\langle \psi, \psi' \rangle = \overline{\psi} \psi' = \psi^{\dagger} \gamma \psi' $ is preserved by $\Spin(p,q)$. The matrix $\gamma$ is hermitian and satisfies

\begin{align}
{S^{ab}}^{\dagger} \gamma = - \gamma S^{ab} \;\; \forall a,b.
\end{align}

In an even number of spacetime dimensions, there are two possible independent solutions for $\gamma$. The matrix $\gamma$ may be proportional to the product of the hermitian gamma matrices $\prod_{a=0}^{p-1} \gamma^a$, or it may be proportional to the product of the antihermitian gamma matrices $\prod_{a=p}^{n-1} \gamma^a$. These two solutions are related up to a factor by multiplication by $\gamma^n \sim \prod_{a=0}^{n-1} \gamma^a$. In an odd number of spacetime dimensions, the matrix $\gamma^n$ is proportional to the identity matrix, and therefore the two solutions are not independent.

As noted in \cite{Pin}, the pin groups $\Pin(p,q)$ and $\Pin(q,p)$, which double cover $\OO(p,q)$ and $\OO(q,p)$ respectively, are not in general isomorphic. In particular, parity transformations on spinors (pinors) are represented as different operators in the two groups. There is also a sign ambiguity in the parity operator. Finally, in an even number of spacetime dimensions, there are two different surjective homomorphisms that map $\Pin(p,q)$ onto $\OO(p,q)$. However, in general a parity transformation $P_i$ which inverts the $i$-th spatial axis is given by either $\pm \gamma_i \prod_{a=0}^{n-1} \gamma^a$ or $\pm \gamma_i$.

In an even number of spacetime dimensions, the two solutions for the inner product matrix $\gamma$, which will be denoted by $\gamma_{\pm}$, result in actions which transform with a $+$ sign and a $-$ sign respectively under parity. Suitable linear combinations of these parity symmetric and parity antisymmetric actions can be taken to obtain a theory of chiral fermions. This is equivalent to using the chirality projection operator to project out the different chiralities.

For convenience, we will construct actions principles that are invariant under $\ISO(n)$/$\SO(n+1)$ that are equivalent to the action \eqref{Dirac action}. However, similar conclusions hold for any other group signature.

\subsection{$\SO(n+1)$ coupling}
Consider the action

\begin{align}
S_F=  \frac{\alpha}{2} \int  &(\D\phi)^A \wedge (\D\phi)^B \ldots \wedge (\D\phi)^X \wedge \left[ \overline{\psi} \gamma^Y \D \psi - (\D \overline{\psi}) \gamma^Y  \psi \right] \epsilon_{AB \ldots XYZ} \phi^Z \label{even fermion action}
\end{align}
in $n$ spacetime dimensions with metric signature $(n,0)$. There are $(n-1)$ instances of $\D\phi$ in this action. The spinor $\psi$ is in a representation of $\Spin(n+1)$. The gamma matrices generate the real Clifford algebra $\mathrm{Cl}(n+1,0)$,

\begin{align}
\{ \gamma^A, \gamma^B \} = 2\delta^{AB}.
\end{align}
The action of the covariant derivative on the spinor $\psi$ is given by

\begin{align}
\D \psi = (\dd + A_{BC} S^{BC}) \psi,
\end{align}
where $A_{BC}$ are the real components of the $\Spin(n+1)$ connection and $S^{BC}= \frac{1}{4} [\gamma^B, \gamma^C]$ are the generators of $\Spin(n+1)$. The spinor inner product is given by $\bah\psi \psi = \psi^{\dagger} \Gamma \psi$, with $\Gamma$ a hermitian matrix satisfying

\begin{align}
{S^{AB}}^{\dagger} \Gamma = - \Gamma S^{AB} \;\; \forall A,B.
\end{align}

For now we assume $n$ is even. In that case the gamma matrices are taken to be $\gamma^Y = \left( \begin{array}{c} 
\gamma^a \\
\gamma^n \end{array} \right)$. The final gamma matrix is defined so that it is proportional to the product of all the others, $\gamma^n \sim \gamma^0 \gamma^1 \ldots \gamma^{n-1}$, with a possible factor of $i$ to ensure that it is hermitian. 

For $n$ even, the representation spaces of $\Spin(n+1)$ and $\Spin(n)$ are isomorphic as vector spaces, and are also isomorphic as inner product spaces upon choosing $\Gamma = \gamma = \gamma_{+}$. Therefore the spinor $\psi$ may equally well be regarded as a spinor of $\Spin(n)$. The action \eqref{even fermion action} may be evaluated in the physical gauge,
 
 \begin{align}
 \phi^A \rightarrow \left( \begin{array}{c} 0 \\
 c \\
 \end{array} \right), 
 \end{align}
whereupon the components of the connection $A_{BC}$ are identified with the spin connection $\omega_{bc}$ and frame field $e_b$ as in \eqref{connection}. Upon setting $c=1$, the action \eqref{even fermion action} is exactly equal to the generalised parity-symmetric Dirac action \eqref{Dirac action}. It is possible to obtain the parity-antisymmetric Dirac action by using the group $\SO(n,1)$ and taking $\Gamma= \gamma = \gamma_{-}$. However, it is not possible to obtain chiral fermions by taking a linear combination of the two.

In an odd number of spacetime dimensions, the representation spaces of $\Spin(n+1)$ and $\Spin(n)$ are in general not isomorphic. However, in the Weyl representation the spinor $\psi$ is the direct sum of Weyl components $\chi_1$, $\chi_2$, each of which is in a representation of $\Spin(n)$,

\begin{align}
 \psi = \left( \begin{array}{c} \chi_1 \\
 \chi_2 \\
 \end{array} \right).
\end{align} 
Therefore for $n$ odd, the matrices $\gamma^Y$ are taken to be in the Weyl representation,

\begin{align}
\gamma^Y = \left( \begin{array}{cc} 0 & \sigma^Y \\
 \sigma^Y & 0 \\
 \end{array} \right),
\end{align}
where $\sigma^Y = \left( \begin{array}{c} \sigma^y \\
 1 \\
 \end{array} \right)$, and $\sigma^y$ generate the Clifford algebra $\mathrm{Cl}(n,0)$,
 
\begin{align}
\{ \sigma^a, \sigma^b \} = 2\delta^{ab}.
\end{align}
The inner product matrix is taken to be

\begin{align}
\Gamma = \left( \begin{array}{cc} 0 & \gamma \\
 \gamma & 0 \\
 \end{array} \right).
\end{align}
Evaluating the action \eqref{even fermion action} in the physical gauge with $c=1$ gives the generalised Dirac action \eqref{Dirac action} for two species of fermions $\chi_1$, $\chi_2$. The gauge invariant constraint $\chi_2=0$ may be imposed to obtain just one fermion species.

In an even number of spacetime dimensions, it is possible to have massive fermions by adding a term proportional to

\begin{align}
\int \vol' \; m \overline{\psi} \psi \label{mass term}
\end{align}
to the action \eqref{even fermion action}. This reduces to a mass term in the physical gauge. In odd dimensions, the fermions are necessarily massless because a mass term for $\psi$ in the Weyl representation would mix the two fermion species $\chi_1$, $\chi_2$.

\subsection{$\ISO(n)$ coupling}

The group $\ISpin(n) = \Spin(n) \rtimes \R^n$ is the semi-direct product of $\Spin(n)$ with the translation group $\R^n$. The spinors in \eqref{even fermion action} are taken to be in a representation of $\ISpin(n)$ given by

\begin{align}
S^{ab} = \frac{1}{4} [\gamma^a, \gamma^b], \quad\quad  t^c = 0,
 \end{align}
where $t^c$ are the generators of $\R^4$. In this representation, the translations act trivially. These generators obey the Lie algebra of $\ISO(n)$,

\begin{align}
[t^a, t^b] &= 0, \\
[ S^{ab}, S^{cd}] &=  \delta^{ad} S^{bc} + \delta^{bc} S^{ad} - \delta^{ac} S^{bd} - \delta^{bd} S^{ac}  , \\
[ S^{ab}, t^c ] &=  \delta^{bc} t^a - \delta^{ac} t^b.
\end{align}

The gamma matrices $\gamma^Y$ are taken to be in the sub-representation \eqref{quotientrep},

 \begin{align}
 \gamma^Y = \left( \begin{array}{c} \gamma^y \\
 0 \\
 \end{array} \right),
 \end{align}
where the final $Y=n$ component is the zero matrix. The covariant derivative is

\begin{align}
D &= d + A^B_{\phantom{B}C} S_B^{\phantom{B} C} \nonumber \\
&= d + \omega^b_{\phantom{b}c} S_b^{\phantom{b} c},
\end{align}
where the connection $A^B_{\phantom{B}C}$ is given by \eqref{iso connection}, and the generators are $S_B^{\phantom{B} C} = \frac{1}{4} [ \gamma_B, \gamma^C ] $.

Evaluating the action in the physical gauge, 

\begin{align}
 \phi^A \rightarrow \left( \begin{array}{c} 0 \\
 c \\
 \end{array} \right),
 \end{align}
and setting $c=1$ gives the generalised Dirac action \eqref{Dirac action}. A mass term can be included by addition of a term of the form \eqref{mass term} to the action \eqref{even fermion action}. There is no qualitative difference between the even and odd-dimensional cases as there was in the $\SO(n+1)$ case.

There are other possibilities for writing an action that is gauge invariant under the group $\ISO(n)$ and that reduces to \eqref{Dirac action} in the physical gauge. In particular, it is possible to have spinors that transform non-trivially under translations, which is achieved in \cite{GN}.

\chapter{Quantisation of $(2+1)$-dimensional gravity}\label{chapter: Quantisation}

In this chapter we explore the quantisation of $(2+1)$-dimensional gravity. There have been several different approaches to this problem \cite{Carlip}. Witten quantised $(2+1)$-dimensional gravity in \cite{Witten, Witten 2}, exploiting the fact that in $2+1$ dimensions gravity is a Chern-Simons theory. On the other hand, the Ponzano-Regge model \cite{PR, PR model, Baez, Freidel2, Smerlak, wedges} is a sum over histories quantisation of $(2+1)$-dimensional gravity with zero cosmological constant. In the first section of this chapter, I review some aspects of the Ponzano-Regge model. Subsequently the gravitational $\ISO(3)$ theory will be quantised using a novel discrete model to define the partition function. This model only uses data associated to the simplices in the triangulation, and not its dual structure. The approach is somewhat tentative and at present requires the spacetime manifold to have trivial topology. The model is then extended to the Lorentzian case, for which the gauge group is ${\ISO}^{+}(2,1)$, the component of ${\ISO}(2,1)$ that is connected to the identity.

\section{The Ponzano-Regge model}\label{PR section}

In this section I will briefly review the Ponzano-Regge model, starting off with a description of the mathematical apparatus that is necessary in its construction.

\definition A $k$-simplex is the smallest convex set containing $k+1$ points embedded in $\R^{k}$.

The notion of a $k$-simplex generalises the triangle, so a $0$-simplex is a point, a $1$-simplex is a line, a $2$-simplex is a triangle, a $3$-simplex is a tetrahedron, etc. The number $k$ is called the dimension of the simplex. $k$-simplexes may contain lower-dimensional simplexes as subsets, which consist of the smallest convex set containing a non-empty subset of the original $(k+1)$ points.

\definition A simplicial complex $C$ is a set of simplexes that satisfies the following conditions:
\begin{itemize}
\item{If a simplex $a$ is in $C$, then all of the simplexes contained in $a$ are also in $C$. }
\item{The intersection of any two simplexes $a$, $b$ in $C$ is either empty or is another simplex that is contained in both $a$ and $b$. }
\end{itemize}
A simplicial $k$-complex is a simplicial complex where the largest dimensions of any simplex is $k$. Given a topological space $X$, a triangulation of $X$ is a simplicial complex $C$, together with a homeomorphism between $C$ and $X$.

Let $\Delta$ be a triangulation of a $3$-manifold $M$. Dual vertices are points placed at the barycentre of each tetrahedron in $\Delta$. Neighbouring dual vertices are connected by dual edges, which are straight lines that puncture the common triangle of their respective tetrahedra. Dual faces are polygons bounded by the cycle of dual edges that `go around' an edge of the triangulation.

Given a graph, a maximal tree is a connected set of edges that contains all vertices in the graph but does not contain any loops. A choice of maximal tree in the graph defined by $\Delta$ will be denoted $T$. A dual maximal tree is a maximal tree in the graph formed by the set of dual vertices and dual edges. Given a triangulation $\Delta$, a choice of dual maximal tree will be denoted $T^*$. A dual face $t^*$ will be said to belong to the maximal tree $T$ if the edge that punctures it is in $T$.

In the Ponzano-Regge model, one has a triangulation $\Delta$ of a $3$-manifold that is decorated with some additional data. To every oriented dual edge $e^*$, there is associated an element of $\SO(3)/\SU(2)$ that will be denoted by $Q_{e^*}$. These variables satisfy $Q_{-e^*}=Q_{e^*}^{-1}$, where $-e^*$ denotes the same dual edge $e^*$ but with the opposite orientation. For every oriented dual face $t^*$ with ones of its vertices arbitrarily chosen as a basepoint, one may form the variable $Q_{t^*}$. This is defined as the product of the elements on the boundary of the dual face $t^*$ starting at the basepoint and continuing in the direction of the orientation, with possible flips in the dual edge orientations to agree with the orientation of the dual face.

Given the above data, the Ponzano-Regge partition function is defined by

\begin{align}
Z_{PR} = \int  \prod_{e^*} \dd Q_{e^*}  \prod_{t^* \notin T} \delta(Q_{t^*}) \label{PR},
\end{align}
where the measure for the integral is the normalised Haar measure on the group $\SO(3)/\SU(2)$, $\int \dd Q_{e^*} = 1$, and the delta-functions are on the group, so that $\int \dd Q_{e^*} \delta(Q_{e^*}) f(Q_{e^*}) = f(I)$, with $I$ the identity element in the group. Due to the fact that $ \dd Q_{e^*} = \dd Q_{e^*}^{-1} $ and $\delta(Q_{t^*})=\delta(Q_{t^*}^{-1})$, the Ponzano-Regge partition function is independent of the orientations of the dual edges and dual faces.

This object only exists for certain topologies of the $3$-manifold $M$. In \cite{PR model}, a cohomological criterion is presented that distinguishes the cases for which \eqref{PR} exists. The issue is that the set of $\delta$-functions is not necessarily an independent set, and therefore the partition function may develop divergences. However, when it does exist, in \cite{PR model} the partition function is shown to be related to the Reidemeister torsion, which is a topological invariant. This proves the triangulation independence of \eqref{PR} in those cases.

The Sciama-Kibble first order form of the gravitational action in $2+1$ dimensions is

\begin{align}
S = \int e^a \wedge F^{bc} \epsilon_{abc}. \label{SK action}
\end{align}
This action is manifestly invariant under $\SO(2,1)$ and has an additional translational symmetry

\begin{align}
e^a \rightarrow e^a + (\D \psi)^a, \label{shift symmetry}
\end{align}
with $\D$ the gauge covariant derivative and $\psi^a$ an arbitrary differentiable function. The invariance of \eqref{SK action} under this transformation can be proved by integrating by parts and then using the Bianchi identity, 

\begin{align}
\D F=0, \label{bianchi identity}
\end{align} 
where the boundary term is assumed to be zero. The action \eqref{SK action} can be linked to an `unregularised' version of the Ponzano-Regge partition function \eqref{PR} using the procedure in \cite{wedges}, which we follow now. This is done using the `wedge' variables, first introduced in \cite{Reisenberger}. 

\begin{figure}[h!]  
\begin{center}
\includegraphics[scale=0.5]{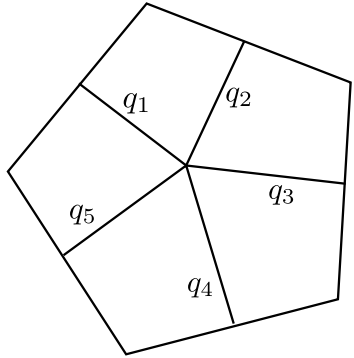}
\end{center} 
\caption{Wedges of a dual face.}\label{wedges fig}
\end{figure}

Wedges are obtained by first placing new vertices at the `centre' of each dual face. This is taken to be the point where the dual face intersects the corresponding edge of the triangulation. Then the centre of each dual vertex is linked to the centre of neighbouring dual faces by line segments. These segments sub-divide each dual face into `wedges'. One can then introduce new variables $X_w$ associated to each wedge $w$. These are elements of the Lie algebra of $\SU(2)/\SO(3)$, and they encode the frame field in \eqref{SK action}. One also introduces $\SU(2)/\SO(3)$ group elements along the oriented segments $s$ which will be denoted by $q_s$. The situation is depicted in figure \ref{wedges fig}.

One may then define the holonomy around an oriented wedge as the product of the group elements around the boundary of the wedge starting at the centre of the dual face and continuing in the direction of the orientation. This holonomy will be denoted $Q_w$. Then the discrete action is defined as

\begin{align}
\hat{S} = \sum_{w} \tr (X_w \log Q_w). \label{wedge action}
\end{align}
This discrete action potentially has an ambiguity if the logarithm function on the group is used na\"\i{}vely. The problem is that the logarithm is multi-valued - there are several possible Lie algebra elements that generate the same group element. This situation regularly occurs in mathematics when, for example, taking the logarithm of a complex number, $y = \ln z$. Consider moving around a circle of radius $1$ in the complex plane; then initially $y(1) = 0$. However, returning to the original point going continuously around the circle one finds that the value of the function changes to $ 2\pi i$.

\begin{figure}[h!]  
\begin{center}
\includegraphics[scale=0.7]{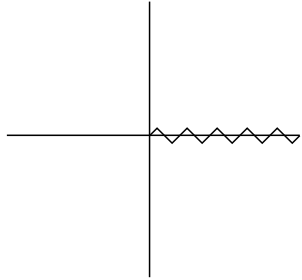}
\end{center} 
\caption{A branch cut in the complex plane.}\label{branch cut}
\end{figure}

The situation is remedied by taking a branch cut, as depicted in figure \ref{branch cut}. This consists of cutting out a line in the complex plane where the function $y=\ln z$ has multiple values. An element of $\su(2)/\so(3)$ may be parameterised by $Z = |Z| n^i \sigma_i$, where $|Z|$ is the length of $Z$, $n^i$ is a unit vector, and $\sigma^i$, $i=1,2,3$ forms a basis. The exponential map covers $\SO(3)$ exactly once if $|Z| \in [0,\pi)$. This region is a sphere of radius $\pi$ centered on the origin of the Lie algebra. Then all the rest of $\so(3)$ is cut out. For $\SU(2)$, the appropriate region is $|Z| \in [0,2\pi)$. The principal domain for the $\SU(2)/\SO(3)$ logarithm will be denoted by $S$, where it is understood by context which region the domain is restricted to in the Lie algebra. From here-on in, all logarithms of group elements will be understood to be the principal logarithm.

The discrete gauge transformations `act at the centre of dual faces'. That is, for each dual face there is an independent gauge transformation whose action is given by

\begin{align}
X_w \rightarrow U X_w U^{-1}, \quad Z_w \rightarrow U Z_w U^{-1},
\end{align}
where $Z_w = \log Q_w$. The action \eqref{wedge action} is invariant under this transformation. After making the appropriate identifications \cite{wedges}, it also has the property that

\begin{align}
\lim_{\Delta \rightarrow M} \hat{S} = S,
\end{align}
where the action on the right hand side is the Sciama-Kibble action \eqref{SK action}, and the limit is the one in which the triangulation $\Delta$ approximates the continuous $3$-manifold $M$.

The partition function is defined as

\begin{align}
Z = \int \prod_{e^*} \dd Q_{e^*} \prod_{s} \dd q_s \prod_w \alpha \dd X_w \; e^{i\hat{S}}. \label{bums}
\end{align}
Here the integral over $X_w$ is the Lebesgue integral over $\R^3$, and $\alpha$ is a normalisation factor which will be specified precisely in what follows. The measure for the $Q_{e^*}$ and $q_s$ integrals is the normalised Haar measure. Performing the $X_w$ integrals yields

\begin{align}
Z = \int \prod_{e^*} \dd Q_{e^*} \prod_{s} \dd q_s \prod_w \alpha (2\pi)^3  \delta^3(Z_w). \label{wedges pf}
\end{align}

Consider $\delta(G)$, the $\delta$-function on the group $\SU(2)/\SO(3)$, with $G=e^Z$ and $Z$ belonging to $S$, the fundamental domain for the principal logarithm on the group. This $\delta$-function is peaked sharply at the identity element and equal to zero everywhere else. It is related to $\delta^3(Z)$ when thought of as a functional on the space of functions on $S$. The defining property of these $\delta$-functions is

\begin{align}
\int \dd G \; \delta(G) f(G) &= f(I), \\
\int_S \dd Z \; \delta^3(Z) g(Z) &= g(0).
\end{align}
The measure for these two integrals are related by

\begin{align}
 dG \;  =  C P^2(Z) \; \dd Z, \label{measures}
\end{align} 
where $P(Z) = \frac{\sin |Z|}{|Z|}$ and $ C = \begin{cases} \frac{1}{2\pi^2} &\mbox{for } \SO(3) \\
\frac{1}{4\pi^2} &\mbox{for } \SU(2) \end{cases}
$.
As a result, the two delta functions are equivalent up to a numerical factor when thought of as functionals on the space of functions on $S$,

\begin{align}
 \delta^3(Z) = C \delta(G). \label{delta relation}
\end{align}
The normalisation factor $\alpha$ in \eqref{bums} is chosen so that 

\begin{align}
\alpha (2\pi)^3 = \frac{1}{C}. 
\end{align}
The partition function \eqref{wedges pf} then becomes

\begin{align}
Z = \int \prod_{e^*} \dd Q_{e^*} \prod_{s} \dd q_s \prod_w  \delta(Q_w).
\end{align}
Next we perform the integrals over the $q_s$ variables. Using the fact that

\begin{align}
\int \dd h \; \delta(g_1 h) \delta(h^{-1}g_2 ) = \delta(g_1 g_2),
\end{align}
we obtain

\begin{align}
Z = \int  \prod_{e^*} \dd Q_{e^*}  \prod_{t^*} \delta(Q_{t^*}).
\end{align}
This is an `unregularised' version of the Ponzano-Regge partition function \eqref{PR}. This partition function is divergent, because the set of delta functions is not independent. The model can be regularised in various ways, one of which is the Turaev-Viro model \cite{TV}. This corresponds to the case of gravity with a non-zero cosmological constant. The wedge variable formalism can be used in the construction of a large number of state sum models, \cite{Witten gauge, wedges, higher BF, Reisenberger2, Reisenberger3, BC, Ooguri, CY, FK, EPR, EPR2, EPRL, Livine, Fairbairn:2006dn}.

One important point to note is that the Ponzano-Regge model in the form \eqref{PR} is a model of `Riemannian quantum gravity', in the sense that one uses the gauge group $\SO(3)/SU(2)$ as opposed to $\SO(2,1)$. One should strictly use the group $\SO(2,1)$ for studying quantum gravity, but this can cause technical difficulties, mainly related to the non-compactness of the gauge group. For an attempt at defining a Lorentzian version of the Ponzano-Regge model, see \cite{Freidel}. `Riemannian quantum gravity' models are typically more tractable than their Lorentzian counterparts, and one hopes that one can still learn something about quantum gravity in this simpler setting.

\section{Quantum $\ISO(3)$ model}
In this section, a new discrete model for the $\ISO(3)$ gravitational theory is constructed. This model only uses data associated to the simplices of the triangulation, and not its dual structure. At present the model is only defined for collapsible triangulations of the $3$-sphere. The model is shown to be equivalent to the Ponzano-Regge partition function \eqref{PR}.

We begin by constructing a discrete analogue of the action \eqref{pagelsaction} in $2+1$ dimensions.

\subsection{Discrete action}
The $\ISO(3)$ gauge gravity action on a closed $3$-manifold $M$ is

\begin{align}\label{3dpagelsaction}
S = \int (\D\phi)^A \wedge {F}^{BC} \epsilon_{ABCD}\, \phi^D, 
\end{align}
where $\phi$ is a multiplet of scalar fields in the vector representation \eqref{vector rep} of the Euclidean group with last component equal to a constant $c$,
\begin{align}\label{lastcomponentc}
\phi=\begin{pmatrix}\phi^1\\\phi^2\\\phi^3\\c\end{pmatrix}.
\end{align} 
This action has an additional local symmetry, besides the $\ISO(3)$ gauge symmetry that is built into the formalism \cite{GN}. This is particular to spacetime dimension three. The shift
\begin{equation}\label{shiftsymmetry}\phi\mapsto\phi+\psi,\end{equation}
 with
\begin{equation}\label{lastcomponent0}\psi=\begin{pmatrix}\psi^1\\\psi^2\\\psi^3\\0\end{pmatrix}\end{equation}
is a symmetry on a closed manifold. To prove this, note that $\Delta S = S(\phi+\psi)-S(\phi)$ has three terms, two of which are immediately zero. The third is  
\begin{align}
\Delta S = \int (D\psi)^A \wedge {F}^{BC} \epsilon_{ABCD}\, \phi^D = c \int (D\psi)^a \wedge {F}^{bc} \epsilon_{abc}\,  ,
\end{align}
where we have noted that ${F}^{BC}$ is zero whenever $B=3$ or $C=3$. This contribution is also equal to zero after integrating by parts and using the Bianchi identity \eqref{bianchi identity}. This means that after gauge fixing the $\phi$ field as in \eqref{physicalgauge}, the action still has an $\ISO(3)$ gauge symmetry that preserves $\phi$, and not just $\SO(3)$ as one would expect from the analysis of the action \eqref{gravityaction2} in a generic dimension. This fact is in accord with the observation that $(2+1)$-dimensional gravity with zero cosmological constant is an $\ISO(3)$ Chern-Simons gauge theory \cite{Witten}.

The discrete action is defined on a triangulation $\Delta$ of $M$ that is decorated with variables as follows:

\begin{itemize}\item For each oriented edge $e$ of the triangulation, an element $Q_e\in\ISO(3)$, consisting of a rotation $M_e \in \SO(3)$ and a translation $b_e \in \R^3$. For the opposite orientation we have $Q_{-e}=Q_e^{-1}$.
\item For each vertex $v$ of the triangulation, a vector $\phi_v\in\R^4$ with last component $c$, as in \eqref{lastcomponentc}. 
\end{itemize}
The edge $e$ has a starting vertex $s(e)$ and a finishing vertex $f(e)$, and the element $Q_e$ is interpreted as the parallel transporter from $s(e)$ to $f(e)$. Edges $e$ and $e'$ satisfying $f(e)=s(e')$ may be composed by concatenation, giving a path consisting of first travelling along the edge $e$ and then $e'$ that is denoted by $e' \, e$. The model is defined by choosing an oriented loop of edges $\gamma_e=e_n\ldots e_2e_1$ that starts and ends at $f(e)$ for each $e$. These edges are arranged so that $f(e_k)=s(e_{k+1})$, which means that the group elements can be composed to give a holonomy for the loop

\begin{align} 
H_e=Q_{e_n}\ldots Q_{e_2}Q_{e_1}.
\end{align}
The choice of the loop $\gamma_e$ for each edge $e$ is given in subsection \ref{collapsible}. For now we note that the choice that is made is such that $\gamma_{-e}= (-e)  (- \gamma_e)e$, where $-e$ denotes the edge $e$ but with opposite orientation. This means that

\begin{align} 
H_{-e}=Q^{-1}_e H_e^{-1} Q_e. 
\end{align}

The discrete version of the integrand  of \eqref{3dpagelsaction} is given by an action for each edge $e$

\begin{align}\label{discreteintegrand}
L_e=\left(\phi_{f(e)}-Q_e\phi_{s(e)}\right)^A\,\left(\log H_e\right)^{BC}\,\epsilon_{ABCD}\,\phi^D_{f(e)},
\end{align}
where the principal logarithm has been used. The first bracket is the obvious discrete version of the $\ISO(3)$ covariant derivative, and the second bracket the analogue of the curvature, with an index raised using $\eta$,

\begin{align} 
\left(\log H_e\right)^{BC}={\left(\log H_e\right)^{B}}_F \eta^{FC}.
\end{align}

Due to the antisymmetry of the epsilon tensor, \eqref{discreteintegrand} can be simplified to

\begin{align}\label{discreteintegrand2}
L_{e}=-\left(Q_e\,\phi_{s(e)}\right)^A\,\left(\log H_e\right)^{BC}\,\epsilon_{ABCD}\,\phi^D_{f(e)}.
\end{align} 
The formula \eqref{discreteintegrand2} has the following important properties:
\begin{itemize}\item It is invariant under $\ISO(3)$ gauge transformations.
\item It is independent of the orientation of the edge $e$.
\end{itemize}
The gauge transformations are the action of the elements $U_v\in\ISO(3)$ independently at each vertex $v$. The group elements $Q_e$ transform as
\begin{align}
 Q_e\mapsto U^{-1}_{f(e)}Q_eU_{s(e)}.
 \end{align} 
The gauge-invariance of \eqref{discreteintegrand2} follows immediately using the fact that 

\begin{align}
\log H_e' &= \log  \left(   U^{-1}_{f(e)} H_e U_{f(e)}  \right) \nonumber\\
&=  U^{-1}_{f(e)} \left( \log H_e  \right) U_{f(e)} .
\end{align}
Reversing the orientation of the edge $e$ gives
\begin{align}\label{reversediscreteintegrand}
L_{-e}&=-\left(Q_e^{-1}\,\phi_{f(e)}\right)^A\,\left(\log H_{-e}\right)^{BC}\,\epsilon_{ABCD}\,\phi^D_{s(e)}\notag\\
&={\left(Q_e^{-1}\right)^A}_E\,\phi^E_{f(e)}\,{\left(Q_e^{-1}\right)^B}_F\,{\left(\log H_e \right)^F}_G\,{Q_e^G}_I\,\eta^{IC}\epsilon_{ABCD}\phi^D_{s(e)}\notag\\
&=\phi^A_{f(e)}\,\left(\log H_e\right)^{BC}\epsilon_{ABCD}\left(Q_e\phi_{s(e)}\right)^D\notag\\
&=L_e.
\end{align}
The discrete action for the whole triangulated manifold is
\begin{align}\label{discreteaction}
S_\Delta= \sum_e L_e.
\end{align}
The sum here is over the set of unoriented edges of  $\Delta$.

This action is $\ISO(3)$ gauge-invariant; however, it does not have the obvious analogue of the additional symmetry  \eqref{shiftsymmetry}. If we choose a vertex $v$ and make the shift $\phi_v\mapsto \phi_v+\psi_v$, with the fields at all other vertices unchanged, the resulting change in the action is

\begin{equation}
\Delta S_\Delta= -\,\psi_v^D\sum_{e\colon f(e)=v} \left(Q_e\phi_{s(e)}\right)^A \left(\log H_e\right)^{BC}\epsilon_{ABCD},
\end{equation}
where the sum is over all oriented edges finishing at $v$. This formula follows by choosing the orientation of each edge $e$ that impinges on $v$ so that $v=f(e)$. This contribution does not vanish in general.

\subsection{Collapsible manifolds} \label{collapsible}

This subsection describes a technical condition on a triangulation called collapsibility that is necessary for the construction of a concrete example of the discrete quantum gravity model.

A collapsing move on a simplicial complex $C$ can occur when there is a $k$-simplex $\sigma$ that is contained in only one $k+1$-simplex $\Sigma$. The move is the removal of both $\sigma$ and $\Sigma$ from the complex \cite{2cxbook}. The complex is said to be collapsible if it can be reduced to a point (a single vertex) by collapsing moves. In fact, if a complex is collapsible, one can always remove the simplexes in dimension order, i.e. in dimension three, remove all 3-2 dimensional pairs first, then 2-1 and finally the 1-0 pairs. 

Consider a triangulation of a closed $3$-manifold $\Delta$ such that removing one tetrahedron $\tau_0$ results in a collapsible complex $C_0=\Delta-\tau_0$. Collapsibility puts very strong constraints on the topology, and in fact $\Delta$ has to be a triangulated three-sphere. 

\begin{figure}[h!]  
\begin{center}
\includegraphics[scale=0.7]{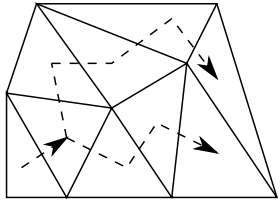}
\end{center} 
\caption{Collapse of a simplicial $2$-complex. The dashed lines form a dual maximal tree.}\label{collapsefig}
\end{figure}

The process of collapse is most easily described using the dual vertices, dual edges, and maximal trees $T$ and $T^*$ that were introduced in section \ref{PR section}. Given a choice of the maximal trees $T$ and $T^*$, first one removes all of the triangles that are punctured by $T^*$, and the interior of all tetrahedra. This can be envisioned as a `burrowing' process, in which one removes $\tau_0$ and then burrows along the paths from $\tau_0$ defined by the dual maximal tree, removing all triangles and the interior of all tetrahedra that are encountered along the way. This is illustrated in two dimensions in figure \ref{collapsefig}.

The set of triangles, edges and vertices that remains after this is a simplicial $2$-complex denoted by $\Delta_{T^*}$.  The middle stage of collapsing is to remove all the edges not contained in $T$ together with all the remaining triangles. In fact the duals of these form a third maximal tree of dual vertices and dual edges in the 2-complex $\Delta_{T^*}$, but this is uniquely determined by $T^*$ and $T$ and is therefore not independent data. After removing all the triangles, all that remains is $T$, which can be collapsed to any one of its vertices.

Notice that in the middle step the collapse sets up a $1-1$ correspondence between edges $e \notin T$ and triangles  $t\notin T^*$. This map assigns to each edge $e \notin T$ the triangle $t_e$ that it collapses through. Since $e$ is in the boundary of $t_e$ there is also a natural correspondence between orientations of $e$ and $t_e$. Therefore the loop $\gamma_e$ that is associated to each edge $e \notin T$ in the discrete Lagrangian \eqref{discreteintegrand} is taken to be the cycle of edges around the boundary of $t_e$. This loop starts and ends at $f(e)$. For $e \in T$, $\gamma_e$ is chosen to be the trivial loop at $f(e)$. The discrete action \eqref{discreteaction} is then

\begin{align}\label{discrete action yo}
S_{\Delta}= \sum_{e\notin T} -\left(Q_e\,\phi_{s(e)}\right)^A\,\left(\log H_e\right)^{BC}\,\epsilon_{ABCD}\,\phi^D_{f(e)}.
\end{align}

\subsection{Partition function}\label{subsec shite}

In principle, one should define the partition function by integrating over the space of all the field variables. However both the gauge group and the space of the variable $\phi$ are non-compact and so integrating over gauge-equivalent configurations will give an infinite result. Thus one integrates over the space of orbits of the gauge group. In practice, this is done by `gauge fixing'; the action is evaluated in a basis in which it explicitly depends on the minimal number of degrees of freedom. One then integrates over only these variables in the partition function.

By applying a suitable $\ISO(3)$ gauge transformation, the $\phi_v$ variables transform as $\phi_v\mapsto U_v\phi_v$, and so can all be taken to a constant vector
\begin{equation} \phi_v= \left( \begin{array}{c} 0 \\
0 \\
0 \\
c \end{array} \right).\end{equation}
The Euclidean group element $Q_e$ consists of a rotation $M_e\in\SO(3)$ and a translation $b_e\in\R^3$. 
 The action  \eqref{discrete action yo} then simplifies to
\begin{align}
S_\Delta=  -c^2 \sum_{e\notin T} b_e^a  Z_e^{bc} \epsilon_{abc} = -c^2 \sum_{e\notin T} b_e \cdot Z_e,
\end{align}
with $Z_e^{bc}=\log H_e^{bc}$. The $\ISO(3)$ element $H_e^{BC}$ consists of a rotation $ G_e^{bc} \in \SO(3)$ and a translation $H_e^{b3} \in \R^3$,

\begin{align}
H_e^{BC} = \left( \begin{array}{cc} G_e^{bc} & H_e^{b3} \\
0 & 1 \end{array} \right).
\end{align}
Thus we have $Z_e^{bc}=\log G_e^{bc} \in \so(3)$. 

With this gauge fixing, the partition function is defined by
 \begin{align}
Z_\Delta &=   \int \left(\prod_e \dd M_{e}\right)  \; \left( \prod_{e \notin T } \alpha   \dd b_e \right) \;e^{-i  c^2\sum_{e \notin T} b_e \cdot Z_{e}} \notag\\
&=  \int \left( \prod_e \dd M_{e}\right)  \prod_{e \notin T }    \;   \frac{ (2\pi)^3 \alpha}{c^2}\delta^3(Z_{e}) . \label{partition function}
\end{align}
Here $\dd M_e$ denotes the Haar measure on $\SO(3)$ normalised so that $\int\dd M_e=1$,  $\dd b_e$ is the Lebesgue measure on $\R^3$ and $\alpha\in\R$ is a constant parameterising the possible normalisations of this measure. The $b_e$ integral for $e \notin T$ naturally covers the whole of $\R^3$ since it is the parameter that describes translations. The $b_e$ variables for $e\in T$ do not appear in the action, so the integrals over these variables are omitted. 

If the constants are defined so that $c^2=4\pi \alpha$, then
\begin{align}\label{ponzanoregge}
Z_\Delta = \int  \prod_{e  } dM_e \prod_{e\notin T} \delta(G_e),
\end{align}
where \eqref{delta relation} has been used. The holonomies $G_e$ appearing in this formula are those around the triangles not in the dual tree $T^*$. Therefore $Z_\Delta$ is equal to the formula for the Ponzano-Regge partition function as defined in \cite{PR model}. The formula has edges and dual edges interchanged but this makes no essential difference. 

The main point to make here is that not only is the partition function $Z_\Delta$ equal to the Ponzano-Regge model, but the local formula for the model is the same. This means also that the expectation values of the observables that are functions of the group elements, as defined in \cite{BH}, are the same in the two models. This gives invariants of the graph of edges of $\Delta$ known as the relativistic spin networks.

In fact, the partition function \eqref{ponzanoregge} has some additional symmetry that can be gauge fixed. The manifest rotational symmetry of the discrete action \eqref{discrete action yo} may be used to set all of the rotation elements on edges in the maximal tree to the identity rotation, $M_e = I \; \forall e \in T$. Then the integral over these variables in the partition function is omitted, resulting in a formula similar to \eqref{ponzanoregge}, 
\begin{align}\label{ponzanoreggefixed}
Z_\Delta = \int  \prod_{e\notin T  } dM_e \prod_{e\notin T} \delta(G_e) .
\end{align}
One may then change variables, giving
\begin{align}\label{ponzanoreggefixed2}
Z_\Delta = \int  \prod_{e\notin T  } dM_e \prod_{e\notin T} \delta(M_e)=1 .
\end{align}
This formula is proved by induction on the number of triangles in the $2$-complex $\Delta_{T^*}$. Recall that this is the $2$-complex that remains after all $3-2$ dimensional simplex pairs have been removed along the dual maximal tree $T^*$ in $\Delta$. In the penultimate step of collapse, where one removes edges and triangles from $\Delta_{T^*}$ along a second dual maximal tree $U^*$, there is a non-empty set of `terminal triangles' that we denote by $\mathcal{T}_{0}$. These are triangles that have two edges that belong to the maximal tree $T$. Therefore, the `burrowing' process described in subsection \ref{collapsible} stops on these triangles. Alternatively, they may be described as the triangles that are at the end of branches of the dual maximal tree $U^{*}$. The subscript $0$ indicates the value of a distance function for triangles. For a given triangle $t$, the value of the distance function is the minimum number of dual edges in $U^{*}$ needed to link the barycentre of $t$ to that of any triangle $t' \in \mathcal{T}_{0}$. Hence the terminal triangles are at a distance $0$. As noted in subsection \ref{collapsible}, the collapse process sets up a $1-1$ correspondence between edges $e \notin T$ and triangles $t\notin T^*$. The action of this map on the set of triangles in $\mathcal{T}_{0}$ gives a corresponding set of edges that will be denoted by $\mathcal{E}_0$.

For all edges $e \in \mathcal{E}_0$, the $\delta$-functions that appear in \eqref{ponzanoreggefixed} reduce to $\delta$-functions on the edge,

\begin{align}
\delta(G_e) = \delta(M_e) \quad \forall e \in \mathcal{E}_{0}.
\end{align}
Now consider the non-empty set of triangles that are at a distance $1$. All of these triangles have two edges that belong to $T \cup \mathcal{E}_0$, and one that does not. The set of edges of triangles at distance $1$ that do not belong to $T \cup \mathcal{E}_0$ is denoted $\mathcal{E}_1$. Then, using the following property of $\delta$-functions,

\begin{align}
\delta(Q) f(Q) = \delta(Q) f(I),
\end{align}
it is clear that 

\begin{align}
\prod_{e \in ( \mathcal{E}_0 \cup \mathcal{E}_1) } \delta(G_e) = \prod_{e \in ( \mathcal{E}_0 \cup \mathcal{E}_1) } \delta(M_e).
\end{align}
Let the maximum value of the distance function for any triangle in $\Delta_{T^*}$ be $d$. Then one can repeat the above process iteratively for triangles at distance $0,1,2,...d$. Taking this process to completion results in the identity

\begin{align}
\prod_{e \notin T} \delta(G_e)  =  \prod_{e \notin T} \delta(M_e).
\end{align}
This proves the change of variables formula used in \eqref{ponzanoreggefixed2}.

\subsection{ Quantum ${\ISO}^{+}(2,1)$ model}

The construction presented in this section is not a model of Lorentzian quantum gravity in three dimensions because the gauge group is ${\ISO}(3)$, and yet there is a crucial factor of $i$ multiplying the exponent of the action in \eqref{partition function}. This situation can be remedied by instead using the group $\ISO^{+}(2,1)$, which is the semi-direct product of the connected piece of $\SO(2,1)$ with the translation group $\R^3$. This gives a Lorentzian version of the previous construction.

The construction is the same as for $\ISO(3)$ except that the group is non-compact. At present the model is again only defined for $3$-manifolds that are homeomorphic to the $3$-sphere. The formula \eqref{discreteintegrand} is the same, using the principal logarithm for ${\ISO}^{+}(2,1)$ that maps ${\ISO}^{+}(2,1)$ group elements onto a suitable fundamental domain in the Lie algebra. This logarithm and the fundamental domain are given explicitly in appendix \ref{appendix1}.

Due to the infinite volume of the gauge group, the analogues of \eqref{partition function} and \eqref{ponzanoregge} are infinite. However it is still possible that suitable observables are finite. Although the relativistic spin networks for $\SO^+(2,1)\times \SO^+(2,1)$ have not been studied, a similar issue arises with the relativistic spin networks for $\SO(3,1)$, which are finite for many graphs \cite{Baez:2001fh,Christensen:2005tr}. The $\ISO(2,1)$ model can be gauge fixed yielding a finite partition
function that is identical to \eqref{ponzanoreggefixed2}.

\chapter{Hamiltonian analysis of $(3+1)$-dimensional gauge gravity}\label{chapter: Hamiltonian}

In this chapter, the Hamiltonian analysis for the $(3+1)$-dimensional gauge gravity theory \eqref{new action} is begun following the procedure set out in Dirac's `Lectures on Quantum Mechanics' \cite{Lectures on QM}. The `na\"{i}ve' Hamiltonian takes a simple form in the language of gauge theory. The Hamiltonian analysis presented here is only at a preliminary stage.

\section{$(3+1)$-dimensional gauge gravity}
We seek to carry out Dirac's programme for the gauge gravity theories described in sections \ref{sec:so(n) gauge gravity} and \ref{sec:iso(n) gauge gravity}  in $n=4$ spacetime dimensions. We will examine the $\ISO(p,q)$ theory first, with $p+q=4$ specifying the group signature. The action is

\begin{align}
S = \int_M & (\D \phi)^A 
\wedge  (\D \phi)^B 
\wedge {F}^{C D} 
\epsilon_{ABCDE} \phi^E. 
\label{new action 2}
\end{align}
We will assume that the $4$-manifold $M$ admits a foliation $M = \Sigma \times \R$, with $\Sigma$ a closed $3$-manifold.

`Spacetime' indices will be denoted by Greek letters, so $\mu, \nu, \ldots = 0,1,2,3$. The restriction of these indices to run over the spatial part only will be denoted by lower case indices $i,j,k \ldots = 1,2,3 $. For simplicity, we take the final $F=4$ component of $\phi^F$ to be $c=1$. Then \eqref{new action 2} may be written as

\begin{align}
S = \int_M \dd^4 x \; (\D_{\mu} \phi)^a 
  (\D_{\nu} \phi)^b 
 {R}^{cd}_{\rho \lambda}
\epsilon_{abcd} \epsilon^{\mu \nu \rho \lambda},
\end{align}
where we have noted that the only non-zero contribution to the action \eqref{new action 2} comes from when the index $F$ is equal to $4$. 

The first step is to divide the action into variables whose time derivative appears and variables whose time derivative does not appear,

\begin{align}
S &= \int_M \dd^3 x \hspace{0.6pt} \dd t \Big[  2 \dot{\phi}^a (D_{i} \phi)^b R^{cd}_{jk } \epsilon_{abcd} \epsilon^{ ijk} + \dot{\omega}^c_{i d} (D_{j} \phi)^a (D_{k} \phi)^b \eta^{de} \epsilon_{abce} \epsilon^{ijk}  \nonumber \\
&+ 2 e_0^a (D_{i} \phi)^b R^{cd}_{jk } \epsilon_{abcd} \epsilon^{ijk} + \omega^a_{o b} \Big( 2 \phi^b (D_{i}\phi)^c R^{de}_{jk } \epsilon_{acde} \epsilon^{ ijk} \nonumber \\
&+ (D_{i}\phi)^c (D_{j}\phi)^d \omega^b_{k e} \eta^{ef} \epsilon_{cdaf} \epsilon^{ijk}
- (D_{i}\phi)^c (D_{j}\phi)^d \omega^e_{k a} \eta^{bf} \epsilon_{cdef} \epsilon^{ijk} \nonumber \\
& \quad \quad \quad \quad + 2( \partial_{i} ( D_{j} \phi)^c) (D_{k}\phi)^d \eta^{be} \epsilon_{cdae} \epsilon^{ijk} \Big)  \Big] \label{separation},
\end{align}
where $\eta^{de} = \diag (\overbrace{1,1, \ldots 1}^\text{p times}, \overbrace{ -1, -1, \ldots -1}^\text{q times}) $. The conjugate momenta are given by

\begin{align}
\pi_a(x) &:= \frac{\partial \mathcal{L}}{\partial \dot{\phi}^a} = 2 (D_{i}\phi(x))^b R^{cd}_{jk}(x)  \epsilon_{abcd} \epsilon^{ ijk}, \label{momentum 1} \\
p^{i d}_c(x) &:= \frac{\partial \mathcal{L}}{\partial \dot{\omega}^c_{i d}} = (D_{j}\phi(x))^a (D_{k}\phi (x) )^b \eta^{de} \epsilon_{abce} \epsilon^{ijk}, \label{momentum 2} \\
p^{0 d}_c(x) &:= \frac{\partial \mathcal{L}}{\partial \dot{\omega}^c_{0 d}} = 0, \label{constraint 1}\\
t_a^{\mu}(x) &:= \frac{\partial \mathcal{L}}{\partial \dot{e}^a_{\mu}} = 0 \label{constraint 2}.
\end{align}
All of these equations define primary holonomic constraints. 

The `na\"{i}ve' Hamiltonian is

\begin{align}
H =\int_{\Sigma} \dd^3 x \Big[ -e_0^a \pi_a -\frac{1}{2} \omega_{0b}^a \Big( \phi^b \pi_a - \phi_a \pi^b + 2(D_{i} p^{i})_a^{\phantom{a}b} \Big) \Big], \label{na\"{i}ve hamiltonian}
\end{align}
where the covariant derivative of $p^{i b}_a(x)$ is given by $(D_{i} p^{i}(x))_a^{\phantom{a}b} = (\partial_{i} p^{i}(x))_a^{\phantom{a}b} + \omega_{i d}^b(x) p_{a}^{i d}(x) - \omega_{i a}^c(x) p_{c}^{i b}(x) $, and we have noticed that this is the form of the final three terms in the square bracket in \eqref{separation}.

The Poisson bracket of two suitable functionals $F$, $G$ on phase space is defined by

\begin{align}
\{ F, G \} = \int_{\Sigma} \dd^3 x \hspace{1pt} \Bigg[ \frac{1}{4} \bigg( \frac{\delta F}{\delta \omega^c_{\mu d}(x)} \frac{\delta G}{\delta p_c^{\mu d}(x)} &- \frac{\delta F}{\delta p_c^{\mu d}(x)} \frac{\delta G}{\delta \omega^c_{\mu d}(x)} \bigg) \nonumber \\
+ \frac{\delta F}{\delta e^a_{\mu}(x)} \frac{\delta G}{\delta t_a^{\mu }(x)} &- \frac{\delta F}{\delta t_a^{\mu }(x)} \frac{\delta G}{\delta e^a_{\mu }(x)} \nonumber \\
+ \frac{\delta F}{\delta \phi^a(x)} \frac{\delta G}{\delta \pi_a(x)} &- \frac{\delta F}{\delta \pi_a(x)} \frac{\delta G}{\delta \phi^a(x)}\Bigg].
\end{align}
This definition can be extended to suitable functions on the phase space in a non-rigorous way by allowing the test `functions' for $F$, $G$ to be distributional. The Poisson brackets for the phase space variables are

\begin{align}
\{ \phi^a(x), \pi_b(y) \} &= \delta^a_b \delta^3(x-y), \\
\{ \omega^{\mu}_{ab} (x), p_{\nu cd} (y) \} &= \frac{1}{2} \left( \eta_{ac} \eta_{bd} -  \eta_{ad} \eta_{bc}  \right)\delta^{\mu}_{\nu} \hspace{0.4pt} \delta^3(x-y), \\
\{ e^a_{\mu}(x), t_b^{\nu}(y) \} &= \delta^a_b \delta^{\nu}_{\mu} \delta^3(x-y),
\end{align}
with all other Poisson brackets equal to zero. 

Adding the primary constraints to the `na\"{i}ve Hamiltonian' \eqref{na\"{i}ve hamiltonian} gives the total Hamiltonian

\begin{align}
H_T = H + \int_{\Sigma}  \dd^3 x \left( u^c_{\phantom{0}d} p^{0 d}_c +  r_{\mu}^a t_a^{\mu} + v^a c_a +  w^c_{i d} \psi^{i d}_c    \right) , \label{hamiltonian 1}
\end{align}
where $u^c_{\phantom{0}d}(x)$, $r_{\Delta}^a(x)$, $v^a(x)$ and $w^c_{\sigma d}(x)$ are arbitrary functions on $\Sigma$, and

\begin{align}
c_a(x) &= \pi_a(x) - 2 (D_{i}\phi(x))^b R^{cd}_{jk}(x)  \epsilon_{abcd} \epsilon^{ ijk}, \label{constraint 4} \\
\psi^{i d}_c(x) &= p^{i d}_c(x) - (D_{j}\phi(x))^a (D_{k}\phi(x) )^b \eta^{de} \epsilon_{abce} \epsilon^{ijk}. \label{constraint 3} 
\end{align}

The non-trivial consistency conditions for the primary constraints \eqref{constraint 1} and \eqref{constraint 2} read

\begin{align}
\dot{t}_a^0(x) &= \{ t_a^0(x), H_T \} \nonumber \\ 
&= \{ t_a^0(x), H \} \nonumber  \\
&= \pi_a(x), \label{pi constraint} \\
\nonumber \\
\dot{p}^{0}_{cd} (x) &= \{ p^{0 }_{cd}(x) , H_T \}\nonumber  \\
&= \{ p^{0}_{cd}(x) , H \} \nonumber  \\
&= \frac{1}{2} J_{cd}(x), \label{j constraint}
\end{align}
where $J_{cd} = \phi_d(x) \pi_c(x) - \phi_c(x) \pi_d + 2(D_{i} p^{i}(x))_{cd}$.

The Poisson bracket of $\omega_{0}^{cd}(x)$ and $e^a_{\mu}(x)$ with the total Hamiltonian fixes the functions $u^{cd}$ and $r_{\mu}^a$,

\begin{align}
\dot{\omega}_{0}^{cd}(x) = \{ \omega_{0}^{cd}(x), H_T \} = u^{cd}(x), \\
\dot{e}^a_{\mu}(x) = \{ e^a_{\mu}(x), H_T \} = r_{\mu}^a(x).
\end{align}

A lengthy calculation gives the following Poisson brackets for the constraints \eqref{pi constraint} and \eqref{j constraint},

\begin{align}
\{ \pi_a(x), \pi_b(y) \} &= 0, \\
\{ J_{ab}(x), J_{cd}(y) \} &= \left[ \eta_{ad} J_{bc}(x) + \eta_{bc} J_{ad}(x) - \eta_{ac} J_{bd}(x) - \eta_{bd} J_{ac}(x) \right] \delta^3(x-y), \\
\{ J_{ab}(x), \pi_c(y) \} &= \left[ \eta_{bc} \pi_a(x) - \eta_{ac}\pi_b(x)\right] \delta^3(x-y).
\end{align}
Up to a delta function, this is the Lie algebra of $\ISO(p,q)$, with $\pi_a(x)$ the generators of translations and $J_{ab}(x)$ the generators of rotations. 

The consistency conditions for the other primary constraints \eqref{constraint 4}, \eqref{constraint 3}, as well as the secondary constraints $J_{ab}(x)$ and $\pi_a(x)$, have not yet been computed. Thus the Hamiltonian analysis presented here is only at a preliminary stage. The full analysis is a challenge for future work.

In the physical gauge, 

\begin{align}
\phi^a(x) \rightarrow 0,
\end{align}
it is possible to show that

\begin{align}
\frac{1}{4} e_0^a(x) \pi_a(x) = -\frac{N^2(x)}{e(x)} h(x) - N^{i}(x) h_{i}(x),
\end{align}
where $e(x)=\det e^a_{\mu}(x)$, and $N(x)$ and $N^{i}(x)$ are called the lapse and shift functions respectively. The functions $h(x)$ and $h_{i}(x)$ are normally called the `Hamiltonian' and `diffeomorphism' constraints respectively, and are given by

\begin{align}
h(x)&= \Pi^{i b}_{a}(x)  \Pi^{j}_{bc}(x) R_{ij }^{ca}(x),\\
h_{i}(x) &= \Pi^{j}_{ab}(x) R_{ i j }^{ba}(x)   ,
\end{align}
with $\Pi^{i}_{ab}(x) = e(x)[e^0_a(x) e^{i}_b(x) - e^0_b(x) e^{i}_a(x)] $.

The expression for the `na\"{i}ve' Hamiltonian is actually completely general for any spacetime dimension $n \geq 4$, and all the calculations that follow \eqref{na\"{i}ve hamiltonian} hold in $n \geq 4$ with the indices generalised appropriately.

The Hamiltonian analysis of the $\SO(p,q)$ theory in spacetime dimension $4 = p+q -1$ proceeds very similarly to the calculation of this section. In particular, the `na\"{i}ve' Hamiltonian is

\begin{align}
H = \frac{1}{2} \int_{\Sigma} \dd^3 x \hspace{0.6pt}   A^B_{0 \hspace{0.6pt} C} J_B^{\phantom{B}C},
\end{align}
where $B,C \ldots = 0 \ldots 4 $, and $A^B_{\mu \hspace{0.6pt} C}(x)$ is the $\SO(p,q)$ connection that comprises both the spin connection $\omega_{\mu \hspace{0.6pt} c}^{b}(x)$ and the frame field  variables $e_{\mu}^b(x)$, as defined by \eqref{connection}. The $J_B^{\phantom{B}C}(x)$ are given by

\begin{align}
J_B^{\phantom{B}C}(x) = \phi^C(x) \pi_B(x) - \phi_B(x) \pi^C(x) + 2( D_{i}p^{i }(x) )_B^{\phantom{B}C},
\end{align}
where $p^{i \hspace{0.6pt} C}_B(x)$ is the momentum conjugate to $A^B_{i \hspace{0.6pt} C}(x)$. Up to a delta function, the Poisson commutation relations for $J_B^{\phantom{B}C}(x)$ are that of the Lie algebra of $\SO(p,q)$. This expression for the `na\"{i}ve' Hamiltonian also holds in spacetime dimensions $n \geq 4$, with the indices generalised appropriately.

It is possible that within the gauge gravity formalism, the Hamiltonian analysis and the constraints may be simpler and more amenable to quantisation. This possibility is left open for future investigation.

\chapter{Conclusion}

In part \Rmnum{1} of this thesis, one-dimensional state sum models for scalar and fermion fields minimally coupled to a gauge field were explored. The models are triangulation independent, and have discrete actions that have the continuum action for fermion and scalar fields minimally coupled to a background gauge field as their continuum limit. The partition functions for the state sum models were shown to be equal to the corresponding functional integral with zeta function regularisation. A precise comparison of the `lattice' and zeta function regularisations was carried out in the fermionic case.

With a particular choice of gauge group, the state sum model for the scalar field on the circle is equivalent to the path integral for the harmonic oscillator. The path integral for the harmonic oscillator is usually constructed as the continuum limit of a state sum that is not triangulation independent. The model presented here in chapter \ref{sec: scalar} has the virtue of triangulation independence, and thus in many ways represents a significant simplification.

The fermionic state sum model of chapter \ref{sec: fermions} has a discrete action with an imaginary-valued part that vanishes in the continuum limit. The partition function of the state sum model is exactly equal to the continuum partition function evaluated with zeta function regularisation. One of the conditions of the Nielsen-Ninomiya theorem on fermion doubling is realness of the discrete action. Thus one of the intriguing questions this work raises is whether these facts are related, and to what extent, if any, fermion doubling might be avoided in a more general setting by the use of complex-valued discrete actions.

The over-arching theme of the first part was to explore the construction of theories of matter within the TQFT framework. The one-dimensional setting provided a simple arena for model construction, but the most interesting direction for future work would be to generalise to higher dimensions.

In chapter \ref{chap: Gauge gravity action}, the gauge theory of gravity that was developed by Pagels in the $\SO(n+1)$ case \cite{Pagels}, and Grignani and Nardelli \cite{GN} in the $\ISO(n)$ case was reviewed, but using a simple matrix formalism for the $\ISO(n)$ theory. In chapter \ref{chapter: Coupling to matter} the coupling of the $\SO(n+1)$ theory to scalar and Yang-Mills fields that was proposed by Ha \cite{Ha} was reviewed, and then generalised to the $\ISO(n)$ case. The coupling of the $\SO(n+1)$ theory to fermions that was proposed by Pagels \cite{Pagels} was reviewed, and it was shown that within this formalism, it is not possible to obtain chiral fermions in an even number of spacetime dimensions. A new and simple coupling of the $\ISO(n)$ theory to fermions was proposed in which the translation generators are represented trivially. The resulting action principle appears to be simpler than that proposed by Grignani and Nardelli \cite{GN}. 

In chapter \ref{chapter: Quantisation}, a new discrete quantum model for the $(2+1)$-dimensional $\ISO(n)$ gauge gravity theory was developed. The construction was carried out for collapsible triangulations of $3$-manifolds with the topological type of the $3$-sphere. In this case, the theory was shown to be equivalent to the Ponzano-Regge model. A possible extension of the model to Lorentzian signature was proposed. An interesting direction for future work would be to explore to what extent the methods introduced in quantising the $(2+1)$-dimensional gauge gravity theory can be applied to other spacetime topologies. Another interesting avenue would be to explore whether suitable observables in the Lorentzian $\ISO^{+}(2,1)$ model are finite. Finally, it would be interesting to explore the quantisation of the $(2+1)$-dimensional gauge gravity theory with non-zero cosmological constant, and whether this has any relation to the Turaev-Viro model \cite{TV}.

In chapter \ref{chapter: Hamiltonian} the initial stage of the Hamiltonian analysis of the $(3+1)$-dimensional gauge gravity theory was undertaken. The `na\"{i}ve' Hamiltonian takes a simple form in the language of gauge theory. The constraint analysis presented here is only at a preliminary stage, and the next part is likely to be significantly more complicated. Nonetheless it is possible that the theory may be more amenable to quantisation, and investigating whether this turns out to be the case is a challenge for future work.

\begin{appendices}

\chapter{Principal logarithm on $\ISO^{+}(2,1)$}\label{appendix1}

The group $\ISO^{+}(2,1)$ consists of elements $(g,b)$, where $g\in \SO^{+}(2,1) $ and $b \in \R^3$ is the translation. The logarithm ambiguity only occurs in the $\SO^{+}(2,1) $ subgroup, so it will be sufficient to define a principal logarithm on $\SO^{+}(2,1) $.

As a differentiable manifold, $\SO^{+}(2,1) $ is the hyperbolic upper half plane with a circle fibre at each point. It is isomorphic to $\mathrm{PSL}(2,\R) = \mathrm{SL}(2,\R)/\{\pm I\} $. The Lie algebra will be denoted $\mathrm{so}^{+}(2,1) $, and it is spanned by abstract algebra elements $X$, $Y$ and $H$ with commutation relations

\begin{align}
[H,X] = 2X \quad [H,Y] = -2Y \quad [X, Y] = H.
\end{align}
A canonical representation is given by

\begin{align}
X = \left( \begin{array}{cc} 0 & 1\\
0 & 0\end{array} \right) \quad Y = \left( \begin{array}{cc} 0 & 0\\
1 & 0\end{array} \right) \quad H = \left( \begin{array}{cc} 1 & 0\\
0 & -1\end{array} \right).
\end{align}
Elements of $\SO^{+}(2,1) $ fall into three conjugacy classes: 

\begin{itemize}
\item{Elliptic} elements are conjugate to $K = e^{\theta(X-Y)} = \left( \begin{array}{cc} \cos \theta & \sin \theta\\
- \sin \theta & \cos \theta \end{array} \right)$.
\item{Hyperbolic} elements are conjugate to $A = e^{tH} = \left( \begin{array}{cc} e^t & 0 \\
0 & e^{-t}\end{array} \right)$.
\item{Parabolic} elements are conjugate to $N = e^{sX} = \left( \begin{array}{cc} 1 & s \\
0 & 1\end{array} \right)$.
\end{itemize}
Using this and the adjoint action of the group on its Lie algebra, it is clear that any element of  $\SO^{+}(2,1) $ may be written as the exponential of some element of the Lie algebra. Therefore the exponential map is onto for the group $\SO^{+}(2,1) $. 

A general element may be written as

\begin{align}
g = e^B = e^{xX+yY+hH}.
\end{align}
The matrix $B$ is given by

\begin{align}
B = \left( \begin{array}{cc} h & x\\
y & -h\end{array} \right).
\end{align}
It has the property that $B^2 = (h^2 + xy) I$, where $I$ is the $2 \times 2$ identity matrix. The exponential map can be evaluated explicitly and gives

\begin{align}
g&= \cosh \psi I + \frac{\sinh \psi}{\psi} B \nonumber  \\
&= \left( \begin{array}{cc} \cosh \psi + \frac{h \sinh \psi}{\psi} & \frac{x\sinh \psi}{\psi}  \\
\frac{y\sinh \psi}{\psi} & \cosh \psi - \frac{h \sinh \psi}{\psi} \end{array} \right),
\end{align}
where $\psi = \sqrt{h^2 + xy}$. It can be checked that this is a matrix with real entries and unit determinant. Due to the fact that $\sinh i\psi = i \sin \psi$ and $\cosh i \psi = \cos \psi$, $g$ can become periodic in some directions in the Lie algebra when $h^2 +xy < 0$.

\begin{figure}[h!]  
\begin{center}
\includegraphics[scale=0.6]{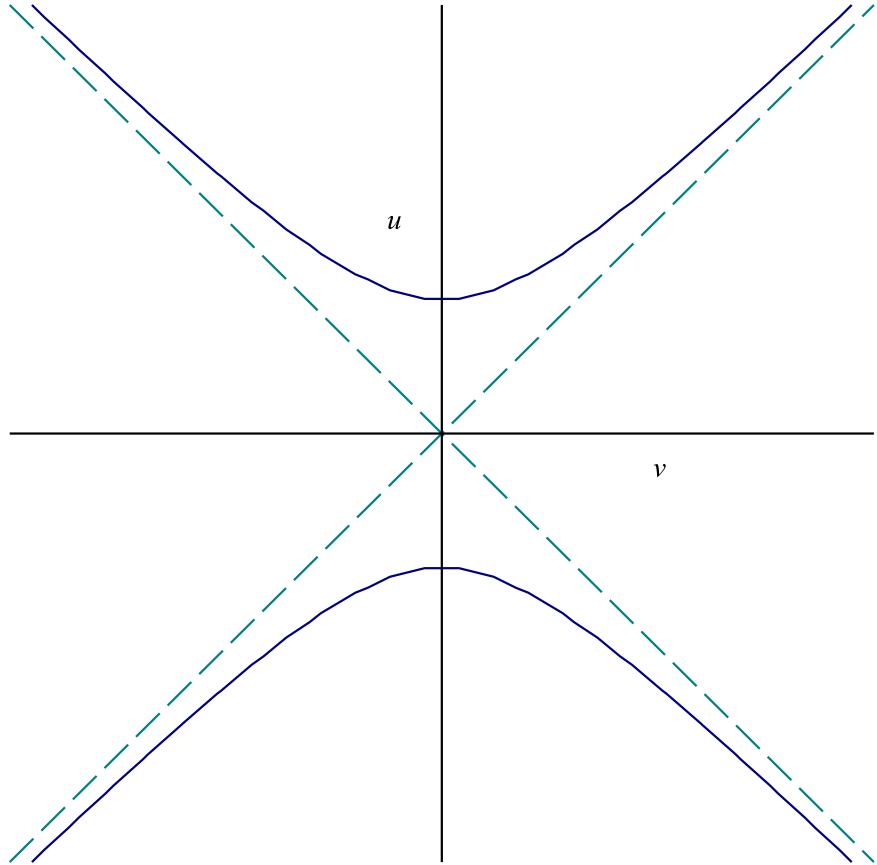}
\end{center} 
\caption{The fundamental domain for the principal logarithm on $\SO^{+}(2,1)$ is given by the interior of the surface of revolution formed by rotating the solid line about the $u$ axis, with all points swept out by the dashed line removed except for the origin. }\label{hyperbola}
\end{figure}

If we define the following coordinates,

\begin{align}
u= \frac{1}{2}(x+y), \nonumber \\
v= \frac{1}{2}(x-y),
\end{align}
then $h^2 +xy = h^2 +u^2 -v^2$. The dividing case $h^2 +u^2 -v^2=0$ is just the equation of a right circular cone that opens on the $v$-axis with aperture $\frac{\pi}{4}$. The region where $\psi$ is imaginary is the interior of this cone. This region is covered conveniently by the following hyperboloid coordinates,

\begin{align}
h = r \sinh w \cos \theta, \nonumber \\
u = r \sinh w \sin \theta, \nonumber \\
v = \pm r \cosh w,
\end{align}
with $w \in [0, \infty)$, $\theta \in [0,2\pi)$. Then $h^2 +u^2 -v^2=r^2$, and in this region the element $g$ is $2\pi$ periodic in $r$. On the other hand, all the Lie algebra elements that lie on the cone are sent to the element $1+B$. Therefore to define a principal domain for the logarithm on $\SO^{+}(2,1) $, we cut out all elements lying on the cone except for the origin. In the interior of the cone, we cut out the region for which $r \notin [-\pi,\pi)$. This is bounded by the hyperboloid $h^2 +u^2 -v^2=\pi^2$. The remaining region is denoted by $S$, and if the exponential map is restricted to this domain, it is $1-1$. The region is depicted in figure \ref{hyperbola}. 

This allows one to unambiguously define the logarithm on $\SO^{+}(2,1)  $, which sends a group element $g$ to the unique Lie algebra element in the domain $S$ that it is the exponential of.

\end{appendices}


\begin{thebibliography}{99}

\bibitem{Kibble}
 T.~W.~B.~Kibble,
  `Lorentz invariance and the gravitational field',
  J.\ Math.\ Phys.\  {\bf 2} (1961) 212.
  
 \bibitem{Sciama}
  D.~W.~Sciama,
  `The Physical structure of general relativity',
  Rev.\ Mod.\ Phys.\  {\bf 36} (1964) 463
   [Erratum-ibid.\  {\bf 36} (1964) 1103].
    
\bibitem{Pop}
 N.~J.~Poplawski,
  `Cosmology with torsion - an alternative to cosmic inflation',
  Phys.\ Lett.\ B {\bf 694} (2010) 181
   [Erratum-ibid.\ B {\bf 701} (2011) 672]
   \href{http://arxiv.org/abs/1007.0587}{[arXiv:1007.0587 [astro-ph.CO]]}.



\bibitem{Sakharov}
 A.~D.~Sakharov,
  `Vacuum quantum fluctuations in curved space and the theory of gravitation',
  Sov.\ Phys.\ Dokl.\  {\bf 12} (1968) 1040
   [Dokl.\ Akad.\ Nauk Ser.\ Fiz.\  {\bf 177} (1967) 70]
   [Sov.\ Phys.\ Usp.\  {\bf 34} (1991) 394]
   [Gen.\ Rel.\ Grav.\  {\bf 32} (2000) 365].


\bibitem{UV gravity}
  M.~H.~Goroff and A.~Sagnotti,
  `The Ultraviolet Behavior of Einstein Gravity',
  Nucl.\ Phys.\ B {\bf 266} (1986) 709.
  
  \bibitem{AS}
  R.~Percacci,
  `Asymptotic Safety',
  In *Oriti, D. (ed.): Approaches to quantum gravity* 111-128
  \href{http://arxiv.org/abs/0709.3851}{[arXiv:0709.3851 [hep-th]]}.

\bibitem{ADM}
 R.~L.~Arnowitt, S.~Deser and C.~W.~Misner,
  `Dynamical Structure and Definition of Energy in General Relativity',
  Phys.\ Rev.\  {\bf 116} (1959) 1322.
  
  
\bibitem{Lectures on QM} P.~Dirac, 
`Lectures on Quantum Mechanics', (Yeshiva Press, New York, 1964).
  
  
  \bibitem{Ashtekar}
 A.~Ashtekar,
  `New Variables for Classical and Quantum Gravity',
  Phys.\ Rev.\ Lett.\  {\bf 57} (1986) 2244.

\bibitem{Baez}
J.~C.~Baez,
  `An Introduction to spin foam models of quantum gravity and BF theory',
  Lect.\ Notes Phys.\  {\bf 543} (2000) 25
  \href{http://arxiv.org/abs/gr-qc/9905087}{[gr-qc/9905087]}.


\bibitem{Connes}
   A.~Connes,
  `Noncommutative geometry and reality',
  J.\ Math.\ Phys.\  {\bf 36} (1995) 6194.


\bibitem{Witten}
E.~Witten,
  `(2+1)-Dimensional Gravity as an Exactly Soluble System',
  Nucl.\ Phys.\ B {\bf 311} (1988) 46.  
  
 \bibitem{circle}
  J.~W.~Barrett, S.~Kerr and J.~Louko,
  `A topological state sum model for fermions on the circle',
  J.\ Phys.\ A {\bf 46} (2013) 185201
  \href{http://arxiv.org/abs/1211.4557}{[arXiv:1211.4557 [math-ph]]}. 
  
  
  \bibitem{Fairbairn:2006dn} 
  W.~J.~Fairbairn,
  `Fermions in three-dimensional spinfoam quantum gravity',
  Gen.\ Rel.\ Grav.\  {\bf 39}, 427 (2007)
  \href{http://arxiv.org/abs/gr-qc/0609040}{[gr-qc/0609040]}.
  
  \bibitem{Dowdall:2010ej} 
  R.~J.~Dowdall and W.~J.~Fairbairn,
  `Observables in 3d spinfoam quantum gravity with fermions',
  Gen.\ Rel.\ Grav.\  {\bf 43}, 1263 (2011)
  \href{http://arxiv.org/abs/1003.1847}{[arXiv:1003.1847 [gr-qc]]}.
  
  \bibitem{Bianchi:2010bn} 
E.~Bianchi, M.~Han, C.~Rovelli, W.~Wieland, E.~Magliaro and C.~Perini,
  `Spinfoam fermions',
  \href{http://arxiv.org/abs/1012.4719}{arXiv:1012.4719 [gr-qc]}.
  
  \bibitem{Han:2011as} 
M.~Han and C.~Rovelli,
`Spinfoam fermions: PCT symmetry, Dirac determinant, 
and correlation functions',
\href{http://arxiv.org/abs/1101.3264}{arXiv:1101.3264 [gr-qc]}.
  
  \bibitem{bolte} 
J.~Bolte and J.~Harrison, 
`Spectral statistics for the Dirac operator on graphs',
J.~Phys. A: Math.\ Gen.\ {\bf 36}, 2747
(2003) 
\href{http://arxiv.org/abs/nlin/0210029}{[arXiv:nlin/0210029]}.

\bibitem{Peskin}
  M.~E.~Peskin and D.~V.~Schroeder,
  `An Introduction to quantum field theory',
  Reading, USA: Addison-Wesley (1995) 842 p

 
 \bibitem{baer-schopka}
C.~B\"{a}r and S.~Schopka, 
`The Dirac determinant of spherical space forms',
in: 
{\it Geometric Analysis and Nonlinear Partial Differential Equations\/}, 
edited by S.~Hildebrandt and H.~Karcher 
(Springer, Heidelberg, 2003) 
\url{http://citeseerx.ist.psu.edu/viewdoc/summary?doi=10.1.1.7.8296}.  


\bibitem{dlmf}
{\it Digital Library of Mathematical Functions\/} 
(National Institute of Standards and Technology, 2011-08-29), 
{\tt \url{http://dlmf.nist.gov/}}.
  

\bibitem{gilkey} 
P.~B.~Gilkey,
{\it Invariance theory, the heat equation, 
and the Atiyah-Singer index theorem\/}, 
2nd edition. 
CRC press (1995). 


\bibitem{paper with mistake}  
G.~V.~Dunne, K.~-M.~Lee and C.~-h.~Lu,
  `On the finite temperature Chern-Simons coefficient',
  Phys.\ Rev.\ Lett.\  {\bf 78} (1997) 3434
  \href{http://arxiv.org/abs/hep-th/9612194}{[hep-th/9612194]}.


\bibitem{vassilevich-manual}
  D.~V.~Vassilevich,
  `Heat kernel expansion: User's manual',
  Phys.\ Rept.\  {\bf 388} (2003) 279
  \href{http://arxiv.org/abs/hep-th/0306138}{[hep-th/0306138]}.

\bibitem{fermion doubling 1}
H.~B.~Nielsen and M.~Ninomiya,
  `Absence of Neutrinos on a Lattice. 1. Proof by Homotopy Theory',
  Nucl.\ Phys.\ B {\bf 185} (1981) 20
   [Erratum-ibid.\ B {\bf 195} (1982) 541].
   
\bibitem{fermion doubling 2}
H.~B.~Nielsen and M.~Ninomiya,
  `Absence of Neutrinos on a Lattice. 2. Intuitive Topological Proof',
  Nucl.\ Phys.\ B {\bf 193} (1981) 173.
  
\bibitem{fermion doubling 3}
 H.~B.~Nielsen and M.~Ninomiya,
  `No Go Theorem for Regularizing Chiral Fermions',
  Phys.\ Lett.\ B {\bf 105} (1981) 219.

\bibitem{HO}
Vadim Kaplunovsky
`Path integral for harmonic oscillator' lecture notes
{\tt \url{http://bolvan.ph.utexas.edu/~vadim/classes/2004f.homeworks/osc.pdf}}. 

\bibitem{bismut-freed}
  J.~M.~Bismut and D.~S.~Freed,
  `The Analysis of Elliptic Families. 2. Dirac Operators, $\eta$ Invariants, and the Holonomy Theorem',
  Commun.\ Math.\ Phys.\  {\bf 107} (1986) 103. 

\bibitem{Feynman}
R.~P.~Feynman and A.~Hibbs,
`Quantum Mechanics and Path Integrals', McGraw Hill, New York,
1965.



\bibitem{Witten 2}
  E.~Witten,
  `Quantum Field Theory and the Jones Polynomial',
  Commun.\ Math.\ Phys.\  {\bf 121} (1989) 351.

\bibitem{MM}
S.~W.~MacDowell and F.~Mansouri,
  `Unified Geometric Theory of Gravity and Supergravity',
  Phys.\ Rev.\ Lett.\  {\bf 38} (1977) 739
   [Erratum-ibid.\  {\bf 38} (1977) 1376].


 
\bibitem{SW}
  K.~S.~Stelle and P.~C.~West,
  `de Sitter gauge invariance and the geometry of the Einstein-Cartan theory',
  J.\ Phys.\ A {\bf 12} (1979) L205.
 
\bibitem{Pagels} 
H.~R.~Pagels,
  `Gravitational Gauge Fields and the Cosmological Constant',
  Phys.\ Rev.\ D {\bf 29} (1984) 1690.
  
  
  
\bibitem{GN} 
G.~Grignani and G.~Nardelli,
  `Gravity and the Poincare group',
  Phys.\ Rev.\ D {\bf 45} (1992) 2719.
  
  \bibitem{Ha}
  Y.~K.~Ha,
  `Coupling of gravity to matter via SO(3,2) gauge fields',
  Gen.\ Rel.\ Grav.\  {\bf 27} (1995) 713
  \href{http://arxiv.org/abs/gr-qc/0409058}{[gr-qc/0409058]}.
  
  \bibitem{Pin}
  M.~Berg, C.~DeWitt-Morette, S.~Gwo and E.~Kramer,
  `The Pin groups in physics: C, P, and T',
  Rev.\ Math.\ Phys.\  {\bf 13} (2001) 953
  \href{http://arxiv.org/abs/math-ph/0012006}{[math-ph/0012006]}.

\bibitem{Carlip}
  S.~Carlip,
  `Quantum gravity in 2+1 dimensions',
  Cambridge, UK: Univ. Pr. (1998) 276 p.


\bibitem{PR}
G.Ponzano and T.Regge, `Semiclassical limit of Racah coefficient' in Spectroscopic and group theoretical methods in physics (Bloch ed.), North-Holland, 1968.


\bibitem{PR model} 
J.~W.~Barrett and I.~Naish-Guzman,
  `The Ponzano-Regge model',
  Class.\ Quant.\ Grav.\  {\bf 26} (2009) 155014
  \href{http://arxiv.org/abs/0803.3319}{[arXiv:0803.3319 [gr-qc]]}.
  
 \bibitem{Freidel2}
  L.~Freidel and D.~Louapre,
  `Diffeomorphisms and spin foam models',
  Nucl.\ Phys.\ B {\bf 662} (2003) 279
  \href{http://arxiv.org/abs/gr-qc/0212001}{[gr-qc/0212001]}.
  
  \bibitem{Smerlak}
  `Bubble divergences from twisted cohomology',
  V.~Bonzom, M.~Smerlak,
  \href{http://arxiv.org/abs/1008.1476}{arXiv:1008.1476 [math-ph]}.
    
    
    \bibitem{wedges}
  L.~Freidel and K.~Krasnov,
  `Spin foam models and the classical action principle',
  Adv.\ Theor.\ Math.\ Phys.\  {\bf 2} (1999) 1183
  \href{http://arxiv.org/abs/hep-th/9807092}{[hep-th/9807092]}.
    
    
    \bibitem{Reisenberger}
  M.~P.~Reisenberger,
  `A Left-handed simplicial action for Euclidean general relativity',
  Class.\ Quant.\ Grav.\  {\bf 14} (1997) 1753
  \href{http://arxiv.org/abs/gr-qc/9609002}{[gr-qc/9609002]}.
  
  \bibitem{Witten gauge}
  E.~Witten,
  `On quantum gauge theories in two-dimensions',
  Commun.\ Math.\ Phys.\  {\bf 141} (1991) 153.
  
  \bibitem{higher BF}
  L.~Freidel, K.~Krasnov and R.~Puzio,
  `BF description of higher dimensional gravity theories',
  Adv.\ Theor.\ Math.\ Phys.\  {\bf 3} (1999) 1289
  \href{http://arxiv.org/abs/hep-th/9901069}{[hep-th/9901069]}.
  
  \bibitem{Reisenberger2}
  M.~P.~Reisenberger and C.~Rovelli,
  `Sum over surfaces' form of loop quantum gravity',
  Phys.\ Rev.\ D {\bf 56} (1997) 3490
  \href{http://arxiv.org/abs/gr-qc/9612035}{[gr-qc/9612035]}.
  
  \bibitem{Reisenberger3}
  M.~P.~Reisenberger,
  `A Lattice world sheet sum for 4-d Euclidean general relativity',
  \href{http://arxiv.org/abs/gr-qc/9711052}{gr-qc/9711052.}
  
  \bibitem{BC}
  J.~W.~Barrett and L.~Crane,
  `Relativistic spin networks and quantum gravity',
  J.\ Math.\ Phys.\  {\bf 39} (1998) 3296
  \href{http://arxiv.org/abs/gr-qc/9709028}{[gr-qc/9709028]}.
  
  \bibitem{Ooguri}
  H.~Ooguri,
  `Topological lattice models in four-dimensions',
  Mod.\ Phys.\ Lett.\ A {\bf 7} (1992) 2799
  \href{http://arxiv.org/abs/hep-th/9205090}{[hep-th/9205090]}.
  
  \bibitem{CY}
  L.~Crane, L.~H.~Kauffman and D.~N.~Yetter,
  `State sum invariants of four manifolds \Rmnum{1}',
 \href{http://arxiv.org/abs/hep-th/9409167}{hep-th/9409167}.
 
 \bibitem{FK}
  L.~Freidel and K.~Krasnov,
  `A New Spin Foam Model for 4d Gravity',
  Class.\ Quant.\ Grav.\  {\bf 25} (2008) 125018
  \href{http://arxiv.org/abs/arXiv:0708.1595}{[arXiv:0708.1595 [gr-qc]]}.
  
  \bibitem{EPR}
  J.~Engle, R.~Pereira and C.~Rovelli,
  `The Loop-quantum-gravity vertex-amplitude',
  Phys.\ Rev.\ Lett.\  {\bf 99} (2007) 161301
  \href{http://arxiv.org/abs/0705.2388}{[arXiv:0705.2388 [gr-qc]]}.
  
  \bibitem{Livine}
  E.~R.~Livine and S.~Speziale,
  `A New spinfoam vertex for quantum gravity',
  Phys.\ Rev.\ D {\bf 76} (2007) 084028
  \href{http://arxiv.org/abs/arXiv:0705.0674}{[arXiv:0705.0674 [gr-qc]]}.
  
  \bibitem{EPR2}
  J.~Engle, R.~Pereira and C.~Rovelli,
  `Flipped spinfoam vertex and loop gravity',
  Nucl.\ Phys.\ B {\bf 798} (2008) 251
  \href{http://arxiv.org/abs/0708.1236}{[arXiv:0708.1236 [gr-qc]]}.
  
  \bibitem{EPRL}
  J.~Engle, E.~Livine, R.~Pereira and C.~Rovelli,
  `LQG vertex with finite Immirzi parameter',
  Nucl.\ Phys.\ B {\bf 799} (2008) 136
  \href{http://arxiv.org/abs/0711.0146}{[arXiv:0711.0146 [gr-qc]]}.
  
  
    
  \bibitem{Freidel}
  L.~Freidel,
  `A Ponzano-Regge model of Lorentzian 3-dimensional gravity',
  Nucl.\ Phys.\ Proc.\ Suppl.\  {\bf 88} (2000) 237
  \href{http://arxiv.org/abs/gr-qc/0102098}{[gr-qc/0102098]}.
  
  \bibitem{Bahr}
  B.~Bahr, B.~Dittrich, F.~Hellmann and W.~Kaminski,
  `Holonomy Spin Foam Models: Definition and Coarse Graining',
  Phys.\ Rev.\ D {\bf 87} (2013) 044048
  \href{http://arxiv.org/abs/1208.3388}{[arXiv:1208.3388 [gr-qc]]}.
 
 \bibitem{2cxbook} `Two-dimensional homotopy and combinatorial group theory', 
edited by Cynthia Hog-Angeloni, Wolfgang Metzler and Allen J. Sieradski, London Mathematical Society Lecture Note Series, 197, Cambridge University Press, Cambridge, 1993.

\bibitem{BH}
  J.~W.~Barrett and F.~Hellmann,
  `Holonomy observables in Ponzano-Regge type state sum models',
  Class.\ Quant.\ Grav.\  {\bf 29} (2012) 045006
  \href{http://arxiv.org/abs/1106.6016}{[arXiv:1106.6016 [gr-qc]]}.



\bibitem{Baez:2001fh}
J.~C.~Baez and J.~W.~Barrett,
  `Integrability for relativistic spin networks',
  Class.\ Quant.\ Grav.\  {\bf 18} (2001) 4683
  \href{http://arxiv.org/abs/gr-qc/0101107}{[gr-qc/0101107]}.

\bibitem{Christensen:2005tr}
   J.~D.~Christensen,
  `Finiteness of Lorentzian 10J symbols and partition functions',
  Class.\ Quant.\ Grav.\  {\bf 23} (2006) 1679
  \href{http://arxiv.org/abs/gr-qc/0512004}{[gr-qc/0512004]}.

\bibitem{TV}
  V.~G.~Turaev and O.~Y.~Viro,
  `State sum invariants of 3 manifolds and quantum 6j symbols',
  Topology {\bf 31} (1992) 865.


\bibitem{Peldan:1993hi}
 P.~Peldan, 
 `Actions for gravity, with generalizations: A Review',
  Class.\ Quant.\ Grav.\  {\bf 11} (1994) 1087
  \href{http://arxiv.org/abs/gr-qc/9305011}{[gr-qc/9305011]}.



\bibitem{QGTQFT}
  J.~W.~Barrett,
  `Quantum gravity as topological quantum field theory',
  J.\ Math.\ Phys.\  {\bf 36} (1995) 6161
  \href{http://arxiv.org/abs/gr-qc/9506070}{[gr-qc/9506070]}. 


\bibitem{Barrett:2013hsa}
  J.~W.~Barrett and S.~Kerr,
  `Gauge gravity and discrete quantum models',
  \href{http://arxiv.org/abs/1309.1660}{arXiv:1309.1660 [gr-qc]}.

\end{thebibliography}
\end{document}